\documentclass[review]{elsarticle}
\journal{arXiv}
\usepackage{amsmath,amsthm,amsfonts}
\usepackage{enumitem}
\usepackage{graphicx}
\usepackage{color}
\usepackage{float}
\usepackage{fixltx2e} 
\usepackage{placeins}
\usepackage{longtable}

\graphicspath{ {./}}
\usepackage{geometry}
\usepackage{lineno,hyperref}
\usepackage{amsmath}
\modulolinenumbers[5]
\newcounter{fig_num} 
\usepackage{mathtools}

\DeclarePairedDelimiter\floor{\lfloor}{\rfloor}

\begin{document}
\begin{frontmatter}
\title{DeePore: a deep learning workflow for rapid and comprehensive characterization of porous materials}

\author[address1]{Arash Rabbani\corref{mycorrespondingauthor}}
\cortext[mycorrespondingauthor]{Corresponding author}
\ead{arash.rabbani@manchester.ac.uk | rabarash@gmail.com}

\author[address1]{Masoud Babaei}
\author[address2]{Reza Shams}
\author[address3]{Ying Da Wang}
\author[address3]{Traiwit Chung}

\address[address1]{The University of Manchester, Department of Chemical Engineering and Analytical Science, Manchester, UK}
\address[address2]{Sharif University of Technology, Chemical and Petroleum Engineering Department, Tehran, Iran}
\address[address3]{School of Minerals and Energy Resources Engineering, University of New South Wales, Sydney, Australia}

\begin{abstract}

DeePore \footnote{GitHub Repository: \url{https://github.com/ArashRabbani/DeePore}} is a deep learning workflow for rapid estimation of a wide range of porous material properties based on the binarized micro--tomography images. By combining naturally occurring porous textures we generated 17700 semi--real 3--D micro--structures of porous geo--materials with size of $256^3$ voxels and 30 physical properties of each sample are calculated using physical simulations on the corresponding pore network models. Next, a designed feed--forward convolutional neural network (CNN) is trained based on the dataset to estimate several morphological, hydraulic, electrical, and mechanical characteristics of the porous material in a fraction of a second. In order to fine--tune the CNN design, we tested 9 different training scenarios and selected the one with the highest average coefficient of determination ($R^2$) equal to 0.885 for 1418 testing samples. Additionally, 3 independent synthetic images as well as 3 realistic tomography images have been tested using the proposed method and results are compared with pore network modelling and experimental data, respectively. Tested absolute permeabilities had around 13 \% relative error compared to the experimental data which is noticeable considering the accuracy of the direct numerical simulation methods such as Lattice Boltzmann and Finite Volume. The workflow is compatible with any physical size of the images due to its dimensionless approach and can be used to characterize large--scale 3--D images by averaging the model outputs for a sliding window that scans the whole geometry. 

\end{abstract}

\end{frontmatter}

\newpage

\section{Introduction}
Data science is becoming an essential tool to analyze the structural features of porous materials based on the tomography images \cite{liu2015machine, kalidindi2015application, steinmetz2016analytics}. The behavior and performance of porous materials are strongly related to the characteristics of its internal micro--structure. In order to discover the descriptive features and the process–structure–property relationships in a porous material, we need to achieve a reliable representation of the internal structure of the porous materials \cite{liu2015machine, niezgoda2011understanding, niezgoda2013novel,fernando2020inter}. Spatial description of such micro--structures have created added--value in diverse fields of studies, from composite material engineering \cite{fast2011new, knezevic2007fast, landi2010multi} and food processing \cite{derossi2017characterizing,derossi2013statistical}, to the petroleum and petrochemical industries \cite{ blunt2013pore, andra2013digital}. For instance, during the past two decades, the field of digital rock physics grew rapidly and showed outstanding advances owing to the power of imaging and analysis techniques \cite{blunt2013pore, blunt2017multiphase}. Based on the captured images, we are able to build realistic simulation models and run many digital measurements and experiments on porous material such as pore and throat sizes, hydraulic and electric conductance, two--phase displacement, and mechanical deformations \cite{huang2016multi,rabbani2020triple}.
Direct calculation of the abovementioned physical properties based on the tomographic data could be a complicated and computationally expensive task especially in the case of large images \cite{mohammadmoradi2016petrophysical,rokhforouz2016numerical,rabbani2019pore}.
In this regard, machine learning approaches can be utilized to make hybrid \cite{rabbani2019hybrid} or full artificially intelligent models \cite{rabbani2017estimation} which are able to reduce the computational costs significantly while maintaining the level of accuracy.  

In this regard, shallow neural networks are powerful tools for modeling moderately complex problems in a timely and efficient manner \cite{gholami2012prediction, russell2016artificial,van2016machine} while they are not very suitable to predict high orders of non--linearity  \cite{bianchini2014complexity}. On the contrary, deep learning models are capable of estimating a highly non--linear behaviour if they are trained on an adequately diversified and large set of input and output data \cite{schmidhuber2015deep,lecun2015deep}.  
Convolutional neural networks (CNNs) as a particular type of deep neural networks can be used for analyzing data with a recognized grid-like topology, similar to image data \cite{lecun1995convolutional, lecun1989generalization,krizhevsky2012imagenet}. A typical CNN uses several filters to extract higher level features from the input data or images and gradually narrows it down to the specified output features \cite{mnih2015human}. CNNs have been mostly used for image segmentation, recognition, classification, and regression \cite{lecun1995convolutional, krizhevsky2012imagenet, he2016deep}. In mathematical terms, convolution is a spatial operation to transform an original function or data into a secondary realization using an operating kernel \cite{keys1981cubic}. Convolution on an input image could lead to generating negative values which are not usually favorable considering the physical meaning of the output layer in that specific problem. At each level of convolution, we can use a down--sampling method such as maximum or average pooling to condense the volume of data without loosing noticeable amount of information \cite{zhou1988computation}. In many cases CNNs can be followed by some fully--connected dense layers of nodes to give more flexibility to the model \cite{girshick2015fast}. CNNs have been used in many recent porous material studies for different purposes including segmentation of porous media images \cite{niu2020digital,karimpouli2019segmentation}, image quality improvement \cite{kamrava2019enhancing}, super resolution, reconstruction \cite{li2018transfer,wang2018super}, classification \cite{baraboshkin2019deep}, and regression \cite{kreyenberg2019velocity,alqahtani2019machine}. Here we briefly describe a background of these applications and narrow the topic down to the specific approach of the present study.   

\subsection{Image improvement and reconstruction}
Considering the multi--scale nature of many of the porous micro--structures, it is necessary to have plenty of details in images while covering a large volume of the object at the same time. In this regard, super resolution techniques powered by CNNs are valuable tools to be trained on pairs of low and high resolution images. There are plenty of recent studies that have presented quantitative methods to obtain a high resolution tomography image of porous material using images with lower spatial resolution \cite{wang2018super, wang2018porous, wang2018ct, wang2019ct, da2019enhancing, da2019super}. As a recent example, Kamrava \textit{et al.} \cite{kamrava2019enhancing} have used a cross--correlation--based simulation to generate an augmented dataset of porous shale images and make a CNN that is able to improve the image quality of similar porous textures.  
 
The resolution enhancement can go further to a level that we are able to generate a detailed realization of the porous material based on the simple input of noise maps through a specific type of CNNs known as Generative Adversarial Networks (GAN) \cite{mosser2018stochastic, mosser2017reconstruction, feng2019accurate,shams2020coupled}. As an example, Mosser \textit{et al.} \cite{mosser2018stochastic} presented a workflow to train a GAN based on the available 3--D tomography images and to reconstruct similar realizations of the original images, while not making an exact copy of them. Then by looking at the hydrodynamic properties of the constructed porous material, they have evaluated the similarity of the realizations. 
  
\subsection{Classification of porous materials}
CNNs are good tools to classify images based on texture, visible elements, or objects \cite{krizhevsky2012imagenet}. This texture recognition has several applications in material and geological sciences to classify or cluster a dataset of porous material images. 
Additionally, some other researchers utilized the CNN framework for recognition of the materials texture \cite{decost2017exploring,baraboshkin2019deep}. For instance, in geoscience, classifying different types of rocks in terms of mineralogy and micro--structure could be a time consuming and biased task if done by hand, while CNNs have widely been used in the past three years to automate these processes in a timely and efficient manner \cite{floriello2019automatic}.

\subsection{Image--based regression models}
Many diverse physical properties of porous materials have been estimated using CNNs in recent years; from thermal to hydraulic and mechanical features \cite{kamrava2020linking,tahmasebi2020machine}. Wei \textit{et al.}\cite{wei2018predicting} proposed a CNN to predict the effective thermal conductivities of composite materials and porous media with more than 0.98 accuracy ($R^2$) on 100 testing image samples while training on 1400 samples. Additionally, permeability and porosity have been heavily investigated through CNNs \cite{wu2018seeing, srisutthiyakorn2016deep,alqahtani2018deep,araya2018deep}. CNNs are able to take both binary or gray--scale images of porous materials to estimate porosity and permeability with an acceptable error. 
Alqahtani \textit{et al.}\cite{alqahtani2019machine} used CNNs to estimate porosity, average pore size and specific surface of the porous rocks based on both types of 2--D tomography images and found that binary images could give a more accurate estimation of porous material characteristics compared to the gray--scale ones. However, the morphology of the binarized images is highly dependent on the thresholding technique and it suffers from the inherent uncertainty \cite{zhang2019challenges}. In another attempt, Cang \textit{et al.}\cite{cang2018improving} designed a CNN for prediction of physical properties of heterogeneous materials and successfully predicted the Young modulus, diffusion and permeability of the porous material with more than 90\% of certainty on their testing data. Recently, Karimpouli and Tahmasebi \cite{karimpouli2019image} developed a CNN model to estimate P-wave and S-wave velocities based on the cross-sectional images of porous material. They were able to estimate these parameters with coefficients of determination around 0.65, and 0.74, respectively. A recent extension of the Karimpouli and Tahmasebi \cite{karimpouli2019image} work is published by Kamrava \textit{et al.} \cite{kamrava2020linking} to investigate the link between the absolute permeability and morphology of the porous materials. They have used a cross--correlation--based simulation technique to augment an image dataset of sandstones and enriched it with hundreds of synthetic and digital images. Then a CNN structure followed by a dense layer is trained to estimate the absolute permeability values that have been obtained by solving the Stokes equation using Avizo commercial software. The range of permeability variations in their work is around one order of magnitude which could be a subject of improvement.

Considering the above-mentioned categories of CNN applications in porous material research, the present study can be considered as an image--based regression model. In order to improve the applicability of the proposed model, a dimensionless and size--independent approach is introduced to calculate porous material features that enables us to analyze images with any spatial resolution.

\section{Methodology}
In this study, we use an augmented set of semi--realistic tomography images of geological porous material to train a convolutional neural network (CNN). The aim of this artificial intelligence model is to predict multiple physical properties of a porous material based on its pore scale images. We refer to this deep learning model of porous material characterization as \textit{DeePore}. In this section, the devised data augmentation technique, assumptions and methods for building the ground truth data, and the utilized deep learning approach are discussed.

\subsection{Input data augmentation}
The original core of the image dataset is composed of 60 real micro-tomography images which their detailed information and corresponding references are available in Appendix A. Considering the fact that it is critical for CNNs to be trained on a large dataset of images, and due to the limited availability of the diverse and realistic tomography data of porous material, data augmentation is required. Several different methods have been used in the literature for augmentation of the training data \cite{shorten2019survey} such as elastic deformation \cite{castro2018elastic}, mixing images \cite{summers2019improved}, cross--correlation--based simulation \cite{kamrava2020linking}, and adversarial reconstruction \cite{gao2018low}. 

\begin{figure}[H]
	\centering\includegraphics[page=\value{fig_num}, width=1\linewidth]{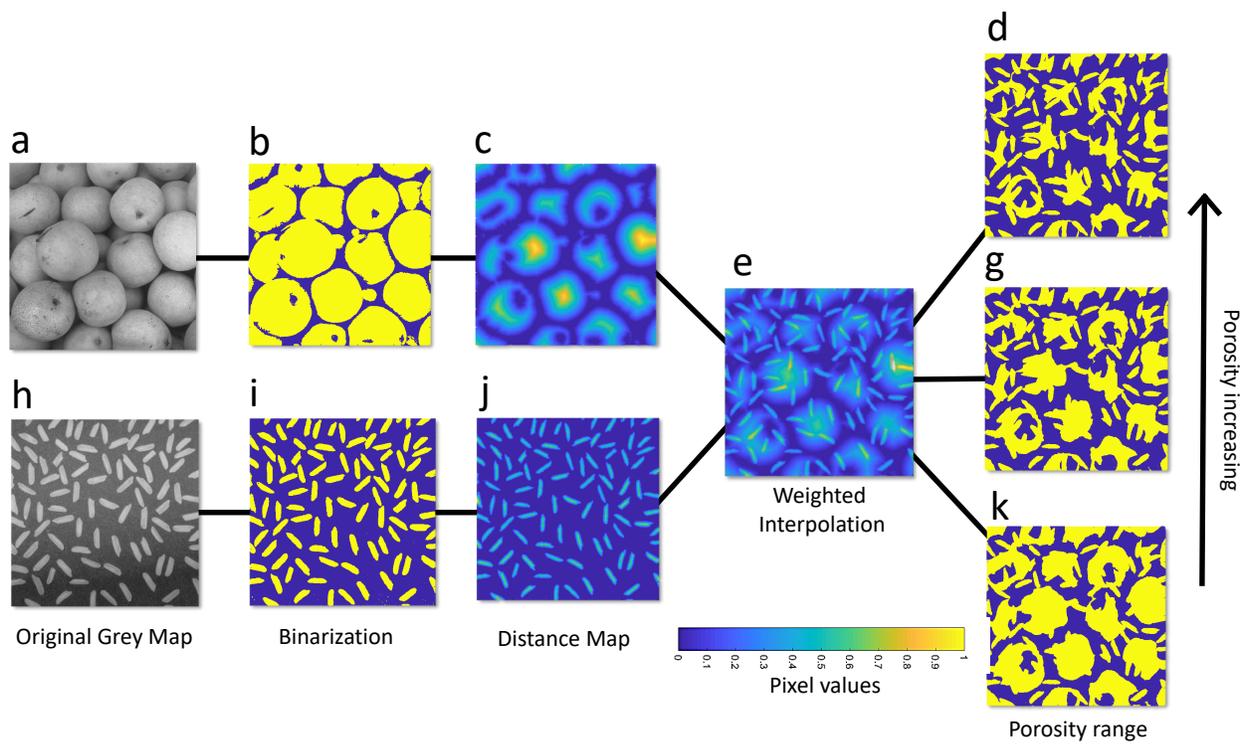} \stepcounter{fig_num}
	\caption{Texture interpolation by weighted averaging of the normalized distance maps, (a and h) original gray maps, (b and i) binarized geometries, (c and j) normalized distance maps of the solid space, (e) equally--weighted average of the distance maps, (d, g and k) three realizations made by changing the threshold level that controls the porosity.}
	\label{fig:interp}
\end{figure}

Inspired by the mixing image method \cite{summers2019improved}, we have adopted a previously developed algorithm to generate more realizations of such data by transforming the existing ones \cite{cohen1998three}. To do this, we select two different images out of the 60 samples and interpolate a hybrid texture among them by weighted averaging of the normalized distance maps. A simplified example of the interpolation technique is illustrated in Fig. \ref{fig:interp}. In this example, two initial grayscale images with different textures (Fig. \ref{fig:interp}-a and h) are binarized using a locally adaptive Otsu algorithm \cite{yan2005multistage} (Fig. \ref{fig:interp}-b and i). Then, normalized maps of the Euclidean distances are calculated (Fig. \ref{fig:interp}-c and j) and combined by weighted averaging to mimic an interpolated texture (Fig. \ref{fig:interp}-e). Then, we can set the threshold level on the obtained hybrid map to reach any desired amount of porosity (Fig. \ref{fig:interp}-d, g and k). 

\begin{figure}[H]
	\centering\includegraphics[page=\value{fig_num}, width=1\linewidth]{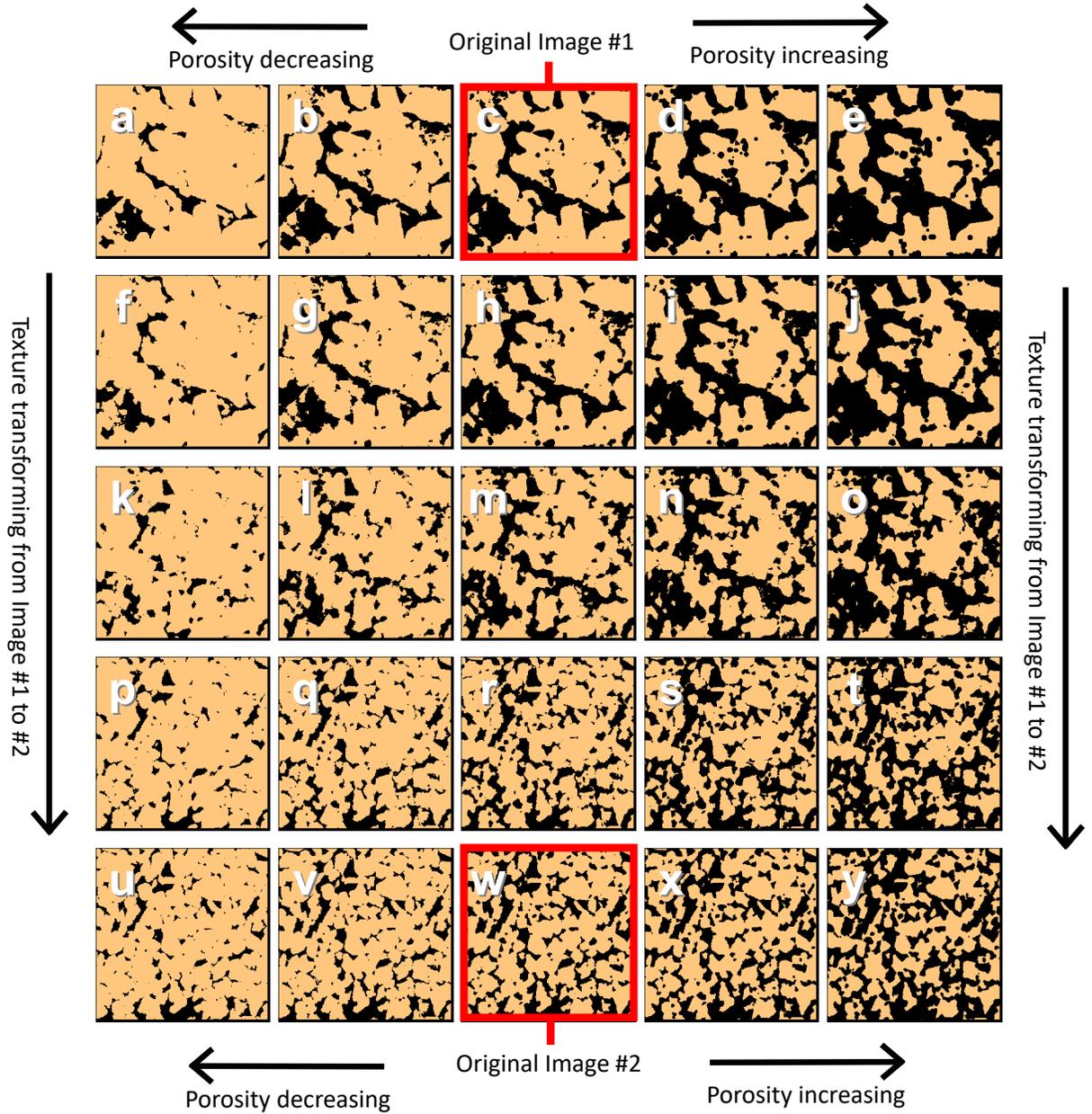} \stepcounter{fig_num}
	\caption{Texture interpolation results based on two real tomography images with a wide range of porosity and texture distributed between two original samples (d and w), porosity of the samples increases from left to right and texture is transforming from c to w when moving downwards.}
	\label{fig:augment}
\end{figure}

As a more realistic example, Fig. \ref{fig:augment} illustrates the texture interpolation results over only two real tomography images (Fig. \ref{fig:augment}-c and w). In this figure, by going from top to the bottom rows, texture is gradually changing from sample \#1 to \#2. Meanwhile, by moving from left to right side of the matrix, porosity is increasing by manipulating the threshold level mentioned above.
 
 It is noteworthy to highlight that the distance maps should be normalized prior to averaging in order to avoid large elements of one image from cloaking the smaller ones in the other image. In order to generate the hybrid realizations of each pair of the real images, we have assumed 10 uniform random numbers between 0 to 1 as interpolation weights, as well as 10 uniform random numbers between 0.1 to 0.45 as final porosity fractions. Then each of the interpolated textures are translated to a randomly selected direction with a random shift within the range of one third of the image width. This random translation process helps to diversify the created dataset and avoid two similar samples which may eventually end up in training and test subgroups and harm the evaluation process by giving a false high performance. After directional shifting of the images, the empty space created is filled by the mirror image of the remaining parts of the micro--structure to maintain the image texture. Based on the data augmentation method described above, the total number of the images in the dataset would be ${60 \choose 2} \times 10$ that yields 17700. However, considering that we aim to calculate several physical properties of these materials, it is expected to filter out outlier geometries with non--physical or null properties that cannot be modelled through the regression techniques. For example, in the case that there is no percolating pathway from one side to the other side of the sample, hydraulic permeability, will be zero and formation factor which indicates electrical resistivity of the void space approaches to infinity. Also, tortuosity will have a null value in such cases. 

\subsection{Building the ground truth data}
\begin{figure}[H]
	\centering\includegraphics[page=\value{fig_num},width=.8\linewidth]{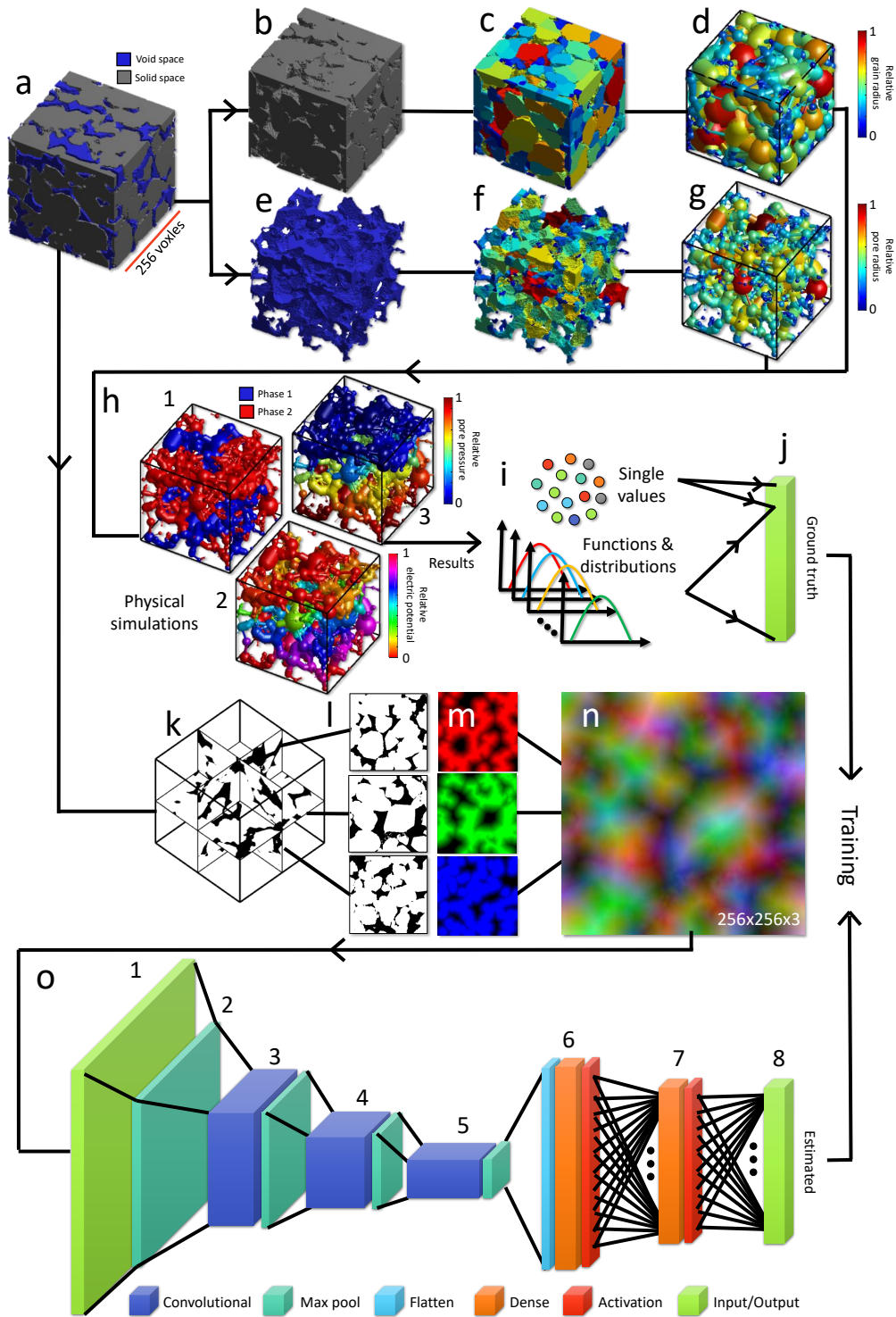} \stepcounter{fig_num}
	\caption{DeePore data workflow for generating the ground truth data and training the CNN based on that, original binary geometry (a), solid and void spaces (b and e), labelled map of nodes (c and f), solid and pore network models (d and g), some of physical simulations on the pore network (h1 to h3), calculated single--value features, as well as the functions and distributions (i), flatten array of ground truth data (j), three perpendicular mid--planes out of the 3--D volumetric data (k), structure of three selected planes with one as solid and zero as void space (l),  three differential distance maps of the solid space which mimics red, green, and blue channels of a synthetic RGB image (Eq. \ref{eq:dmap}) (m), input feature map of the porous media as a fictitious RGB image (n), (o) structure of the designed CNN with 8 layers each of which described in Table \ref{tab:CNN}.}
	\label{fig:data_workflow}
\end{figure}

After construction of the augmented set of image data, we use a series of in--house codes developed based on the available literature to analyze the micro--structures of the porous material. In this regard, pore network modeling (PNM) techniques \cite{xiong2016review} are used to simulate different physics and processes on the 3--D porous samples and the results obtained are assumed to be the ground truth for training the DeePore CNN. In addition to the pore space characterization, we  analyze the solid portion of the images which are defined as a solid network model to enable studying its mechanical behaviour similar to approach presented by Herman in 2013 \cite{herman2013shear} but with 3 dimensions. A solid network model describes the contact area, direction and length between different grains of a porous structure and assimilate the porous structure to a truss which makes it possible to solve finite element equations in a coarse grid model and obtain some mechanical properties of the micro--structures\cite{cortes2019geometry,herman2013shear}.  
A sample realization of the solid and pore networks of a porous material are visualized in Figure \ref{fig:data_workflow}-d and g. In order to construct these networks, initially we need a 3--D binarized image (Fig. \ref{fig:data_workflow}-a) that is segmented to the void and solid spaces which are respectively shown in Figure \ref{fig:data_workflow}-b and e. Then using watershed segmentation algorithm we break down an interconnected micro--structure into a separately labelled map of nodes as can be seen in Figure \ref{fig:data_workflow}-c and f. The color gradient in these illustrations indicates the relative equivalent radius of the nodes extracted for both void (pore) and solid networks. When we have detected location and boundaries of each node, then by analyzing the node map connectivities, two networks can be extracted for both void and solid spaces (Fig. \ref{fig:data_workflow}-d and g). Watershed segmentation algorithm which is used to break down the micro-structures into a mathematically describable 3--D network, has been widely employed for porous material characterization from tomography images \cite{wildenschild2013x,gostick2017versatile,Rabbani2014}. This algorithm uses the Euclidean distance transform of a binary object to detect the narrowest parts of the connections between different nodes. More details on the methodology and validation of watershed segmentation algorithm can be found in \cite{gostick2017versatile,Rabbani2014}.     

Now in order to build the ground truth data for training DeePore CNN, we investigate the constructed pore and solid networks by measuring several morphological features and running physical simulations (Fig. \ref{fig:data_workflow}-h). In this section, we briefly describe the simulation techniques employed and some assumptions made to generalize the analysis of outcomes. As an example, we have illustrated 3 simulation results on a sample pore network model in Fig. \ref{fig:data_workflow}-g. These simulation results are fluid saturation in a 2--phase drainage process, electricity flow through the saturated pore--space, and pore pressure of the single--phase fluid flow, respectively depicted in Fig. \ref{fig:data_workflow}-h1 to h3 to give some insight on the ground truth generation. 

The list of the physical properties and features that have been obtained for each of the samples within the dataset is provided in Table \ref{tab:outputs}. As can be seen, we report 15 single--value features that comprehensively describe the morphological, hydraulic, mechanical and electrical properties of porous material. Additionally, 4 functions and 11 distribution curves are extracted for each porous sample to describe its characteristics (Fig. \ref{fig:data_workflow}-i). Here is a brief introduction to the calculated set of properties and more details regarding the methodology of extracting each of the features is available in the corresponding references in Table \ref{tab:outputs}.

\subsubsection{Morphological properties}
Based on the extracted network models, probability distribution in addition to the average values are reported for pore body radius, pore throat radius, throat length, grain radius, and pore connectivity \cite{Rabbani2014} which is also known as the network coordination number \cite{blunt2013pore}. In addition, pore density that indicates the number of pores per unit volume of the geometry, grain sphericity \cite{beard1973influence}, and specific surface are calculated for each of the 3--D images in the dataset. Furthermore, we have used Dijkstra's algorithm \cite{shanti2014x,dijkstra1959note} to find the shortest path from one face to the other face of the pore networks and calculate tortuosity. In this regard, the shortest path between each two random pairs of the pores from inlet to the outlet of the pore network is calculated for several times and average value of all shortest paths is reported as tortuosity. Finally, as a morphological property of porous material we have calculated the two--point correlation function of the binarized images which shows how well--correlated are the porosity of two random points selected with a specified distance between them \cite{blair1996using}. For more details regarding the calculation methodology for each of the properties please refer to the references provided in Table \ref{tab:outputs}.

\subsubsection{Hydraulic properties}
Absolute and relative permeabilities are calculated based on the extracted pore networks. For calculating the fluid conductance in each pore throat, realistic cross--sectional shapes of the throats are used to provide better results \cite{rabbani2019hybrid}. Additionally, for calculation of the two--phase flow functions, we have assumed zero contact angle in the case that two immiscible fluids are present within the porous media. Also, smooth spherical curvatures are assumed for pores and throats to simplify the displacement process. A thin layer of the wetting phase fluid is present at the wall surface to maintain the phase connectivity but it does not contribute to the hydraulic conductivity of the throats. The procedure we use to model two--phase displacement in a pore network is a quasi--static approach with stepwise increment of the non--wetting phase pressure and domination of the capillary forces over the viscous forces. The quasi--static approach is fully described by Valvatne and Blunt \cite{valvatne2004predictive}. The capillary pressure curve is another important hydraulic property of the porous material and as discussed, we require a technique to remove the pressure unit of this parameter. To this end, we have used the concept of Leverett J-function curve \cite{chaouche1994capillary} which is a dimensionless version of the capillary pressure normalized for different porosities, permeabilities, contact angle and interfacial tension between the two displacing fluids in porous media. 

\subsubsection{Electrical properties}
Formation and cementation factors are two electrical properties of porous material that have been calculated using the extracted pore network models \cite{oren1998extending,man2000pore}. These parameters are critical in Archie's equation \cite{archie1942electrical} and helps to describe the electrical behaviour of a porous medium saturated with a conductive fluid. Formation factor is the ratio of the electrical resistance of the fully saturated porous media to the electrical resistance of the pure fluid \cite{adler1992formation}. In order to calculate this feature, we solve a resistor network assuming conductive fluid inside the pore space using finite difference method. Also, with a similar approach, this value remains the same when we are measuring the ratio of the mass diffusivity of a component in a bulk fluid relative to its diffusivity through the fully--saturated porous media \cite{sevostianov2017connection,aminnaji2019effects}.   

\subsubsection{Mechanical property}
In addition to many pore--dependent properties, we have modelled relative Young modulus of the material which is a solid phase feature. For this purpose, we assume that the extracted solid network is a truss--like structure and by applying normal compressional force on each side of the geometry, the directional Young modulus is calculated by dividing the normal stress over the strain ratio \cite{buxton2007bending,wallach2001mechanical}. Then arithmetic average of the directional values is calculated and divided by the Young modulus of the pure non--porous material to obtain the relative Young modulus which is a dimensionless number \cite{gor2015elastic}.

\subsubsection{Dimensionless approach}
It is noteworthy that we have removed the original spatial resolution of the data and defined a unified unit of length which is equal to the physical size of each voxel. For example, the unit of absolute permeability which is area has become $px^2$ which means that we need to multiply the resulted permeability by the spatial resolution to the power of two in order to retrieve the re--scaled permeability value. Here, we used $px$ as a short form for pixel or voxel size which is our unit of length. In the same manner, all other reported features are dimensionless or described only in length unit which is convertible to voxel size. The list of all alternative units is presented in  Table \ref{tab:outputs}. For features that can be calculated directionally such as permeability, we have assumed an isotropic structure and reported the arithmetic average of the values in \textit{x}, \textit{y}, and \textit{z} directions. Although in some cases this averaging does not have explicit physical meaning, but it is used to cover directional non--conformities in the porous structures that can affect the extensibility of the model.

\begin{table}[H]
	\footnotesize
	\centering
	\begin{tabular}{c|c| c |c |c}
		\hline	
		Num. & Output indices & Feature & Data type & Reference\\
		\hline
		1 & 1 & Absolute permeability ($px^2$)  & Single value & \cite{rabbani2019hybrid}\\
		2 & 2 & Formation factor (ratio) & Single value & \cite{oren1998extending}\\
		3 & 3 & Cementation factor (ratio) & Single value & \cite{man2000pore}\\
		4 & 4 & Pore density ($1/px^3$) & Single value & \cite{cooper2004comparison}\\
		5 & 5 & Tortuosity (ratio) & Single value & \cite{shanti2014x,dijkstra1959note,menon2020pore}\\
		6 & 6& Average coordination number & Single value & \cite{Rabbani2014}\\
		7 & 7 & Average throat radius ($px$) & Single value & \cite{rabbani2016estimation}\\
		8 & 8 & Average pore radius ($px$) & Single value & \cite{rabbani2016estimation}\\
		9 & 9 & Average throat length ($px$) & Single value & \cite{rabbani2016estimation}\\
		10 & 10 & Average pore inscribed radius ($px$) & Single value & \cite{jiang2007efficient} \\
		11 & 11 & Specific surface ($1/px$) & Single value & \cite{rabbani2014determination,rabbani2014specific}\\
		12 & 12 & Average throat inscribed radius ($px$) & Single value & \cite{jiang2007efficient}\\
		13 & 13 & Grain sphericity (ratio) & Single value & \cite{beard1973influence}\\
		14 &14 & Average grain radius (ratio) & Single value & \cite{rabbani2015comparing}\\
		15 & 15 & Relative Young module (ratio) & Single value & \cite{buxton2007bending,wallach2001mechanical,yeheskel2000new}\\
		16 &16-115 & Leverett J--function (ratio) & Function & \cite{chaouche1994capillary}\\
		17 &116 - 215 & Wetting relative permeability (fraction) & Function & \cite{valvatne2004predictive}\\
		18 & 216 - 315 & Non-Wetting relative permeability (fraction) & Function & \cite{valvatne2004predictive} \\
		19 & 316 - 415 & Two-point correlation function ($1/px$) & Function & \cite{blair1996using}\\
		20 & 416 - 515 & Pore radius distribution ($px$) & Distribution & \cite{rabbani2017evolution,baychev2019reliability}\\
		21 & 516 - 615 & Throat radius distribution ($px$) & Distribution & \cite{rabbani2017evolution,baychev2019reliability}\\
		22 & 616 - 715 & Throat length distribution ($px$) & Distribution & \cite{rabbani2017evolution,baychev2019reliability}\\
		23 & 716 - 815 & Pore inscribed radius distribution ($px$) & Distribution & \cite{jiang2007efficient}\\
		24 & 816 - 915 & Throat inscribed radius distribution ($px$) & Distribution & \cite{jiang2007efficient}\\
		25 & 916 - 1015 & Throat average distance ($px$) & Distribution & \cite{rabbani2019hybrid}\\
		26 & 1016 - 1115 & Throat permeability distribution ($px^2$) & Distribution & \cite{rabbani2019hybrid}\\
		27 & 1116 - 1215 & Coordination number distribution  & Distribution & \cite{Rabbani2014}\\
		28 & 1216 - 1315 & Pore sphericity distribution (ratio) & Distribution & \cite{kong2018pore}\\
		29 & 1316 - 1415 & Grain sphericity distribution (ratio) & Distribution & \cite{beard1973influence}\\
	30 & 1416 - 1515 & Grain radius distribution ($px$) & distribution & \cite{rabbani2015comparing}\\
		\hline
\end{tabular}
\smallskip
\caption{List of the physical features of porous material which are considered to be the outputs of the model in addition to the corresponding units and references that describe the methodologies in detail.}
\label{tab:outputs}
\end{table}

\subsection{Deep learning method}
We have generated the dataset of semi--realistic micro--structures of porous material and a wide range of 30 physical properties are calculated for each of the samples. Now we aim to build a machine learning model that is able to estimate these properties purely by analyzing input images and learning an implicit knowledge of the underlying physics. It should be noted that this \textit{implicit} knowledge is different than the physics-informed machine learning models that \textit{explicitly} embed physical equations in the structure of their network layers \cite{raissi2017physics}.   

As discussed, CNNs have proved to be efficient in image classification, segmentation, and regression. So, we have designed a CNN structure combined with two dense layers of neurons to make a regression model that is able to estimate all physical properties of porous material mentioned above in a supervised manner. Data workflow and CNN structure are presented in Fig. \ref{fig:data_workflow} and Table \ref{tab:CNN}. Here, we are providing more details regarding the structure of the network and the training process.

\subsubsection{Network input layer}
Initially, we take a 3--D image from the dataset with the size of $256^3$ voxels and extract three perpendicular mid--planes of the volumetric data (Fig. \ref{fig:data_workflow}-k to l). Then, the distance transform of the solid and void spaces is calculated for each of the images and is deduced to make an initial feature map ($M$) as follows:

\begin{equation}
\label{eq:dmap}
M=\floor*{\frac{8}{S}f_d(1-A)-\frac{8}{S}f_d(A)}
\end{equation}

Where $S$ is side size of the image which is 256 voxels in our case, and $\floor*{x}$ is the floor operator that rounds down the decimal points to the closest smaller integer, $f_d$ is the Euclidean distance transform and $A$ is the a 2--D plane cut through the 3--D volume perpendicular to one of the major axes (Fig. \ref{fig:data_workflow}-k). As a matter of fact, variable $A$ is an array that contains 0 representing pore space and 1 for solid voxels. The reason to multiply distance maps by the ratio of $\frac{8}{S}$ is to ensure all the calculated values will be mainly between $-1$ and $1$ which is suitable to be used as CNN input. The distance maps are not only able to deliver information about the original binary map, but also, describe the Euclidean distances between each point of that binary map to the nearest boundary. This additional information enriches the model input layer with more data compared to passing a simple binary array. 

Now, based on the three maps generated using Eq. \ref{eq:dmap}, we generate a fictitious RGB image by stacking them into each other and make an initial feature map to be used as the input for the CNN (Fig. \ref{fig:data_workflow}-n). The term RGB refers to a color space for image quantization composed of three channels of red, green, and blue. The reason to mimic an RGB image is the common use of these image formats as input of a CNN. In addition, RGB images are easy to store and read from hard disk and there are plenty of lossless compression methods invented to minimize their size when stored on disk \cite{gormish1997lossless}. Use of the whole 3--D data as the input of the CNN instead of the perpendicular mid--planes could increase the accuracy of the results, while it can significantly increase the computational expenses which are not desirable. Also this defeats the purpose of this research to propose an efficient while adequately accurate model. 

\subsubsection{Network hidden layers}
At the first layer of CNN, we initially run a 2 by 2 max--pooling filter to reduce the size of the input data without losing too much information (Fig. \ref{fig:data_workflow}-o2). Then, 3 convolutional layers are designed to gradually decrease the size of the information while maintaining the main geometrical features by applying different sequential filters on the input images (Fig. \ref{fig:data_workflow}-o3 to o5). Each convolutional layer is followed by a $2 \times 2$ max--pooling filter to finally make data small enough to be fitted into a fully--connected dense layer. The first dense layer is activated by ReLU, while the second one uses sigmoid (Fig. \ref{fig:data_workflow}-o6 and o7). This network architecture is designed by  testing a range of different structures and monitoring the performance of each training scenario in terms of accuracy. The selected training scenario has three convolution kernels with the size of $3 \times 3$ with stride equal to $1 \times 1$. More details on the structure of the proposed CNN is provided in Table \ref{tab:CNN}.

\subsubsection{Network output layer}
As it can be seen in Table \ref{tab:CNN}, the output layer of the network is a one dimensional array of 1515 elements (Fig. \ref{fig:data_workflow}-o8). The first 15 elements of the array are 15 single--value features calculated for each of the porous samples as described in Table \ref{tab:outputs}, rows 1 to 15. The next 1500 elements of the output array describe 4 functions and 11 distribution curves each of which occupies 100 elements of the array. The range of the array indices for each of the output parameters is described in Table \ref{tab:outputs}.
In order to fit the wide range of variables and functions into an array of 1515 elements, certain reshaping and interpolation operations are required for the raw results of pore scale modelling. The four functions that occupy indices from 16 to 415, are Leverett J-function, wetting relative permeability, non--wetting relative permeability and two--point correlation function. The first three are functions of wetting phase saturation which is a fraction between 0 to 1. So, in order to summarize each of these three curves into 100 elements, we have divided the whole curve into 100 pieces each of which with 0.01 distance from each other in terms of wetting phase saturation. Similarly, for two--point correlation function, we have split the curve into 100 segments each of which with 0.5 voxel distance to the next one, in order to cover a total lag distance of 50 voxels in 100 elements. For more details regarding this function please refer to \cite{blair1996using}. Next, in order to fit each of the 11 distribution curves into 100 elements, we are using the cumulative format of the probability distributions (CDF) that pack a full range of  variable changes into a sigmoid--like curve between 0 and 1. Consequently, we have divided the CDF curve into 100 pieces with 0.01 distance between every two consecutive points in \textit{Y} axis and embedded the corresponding values of \textit{X} axis into the output array. For more clarification, a sample set of the described functions and distribution curves will be presented in the Results and Discussion section (Table \ref{tab:sample_output} and Fig. \ref{fig:sample_output}). 

\subsubsection{Model training}
Development and training of the DeePore CNN is implemented in Python using Keras with TensorFlow backend \cite{abadi2016tensorflow}. Additionally SciPy, Numpy and Matplotlib \cite{hunter2007matplotlib} as open--source packages of Python are used for data pre-- and post--processing. Back--Propagation RMSprop algorithm \cite{riedmiller1993direct} with the learning rate of $10^{-5}$ is used for training the CNN by minimizing the prediction loss in terms of mean squared error. We have used 80 \% of the data samples for training the network, 10 \% for validation and 10 \% are kept outside of the workflow for independent and un--biased testing of the results obtained. We have trained the model for 100 epochs with batch size of 100 samples per each updating of the model gradient. The input and output data are fed into the model using large size Hierarchical Data Format (HDF) files. A Python Generator method, reads the data batch by batch from the HDF file to avoid occupying a large amount of machine memory. 
 
\begin{table}[H]
	\footnotesize
	\centering
	\begin{tabular}{c| c| c| c| c|c}
   \hline
    Layers & Type  & Input size & Kernel & Options & Trainable parameters \\
    \hline
1     & Input & 256$\times$256$\times$3 & -     & Normalization & 0 \\
2     & Pooling & 128$\times$128$\times$3 & -     & Max Pool 2$\times$2 &0 \\
3     & Convolutional & 64$\times$64$\times$6 & 3$\times$3   & Max Pool 2$\times$2 & 336 \\
4     & Convolutional & 32$\times$32$\times$12 & 3$\times$3   & Max Pool 2$\times$2 & 2616 \\
5     & Convolutional & 16$\times$16$\times$18 & 3$\times$3   & Max Pool 2$\times$2 &  7812\\
6     & Fully-connected & 1$\times$1$\times$9217 & -     & ReLU Activation  & 9217$\times$1515 \\
7     & Fully-connected & 1$\times$1$\times$1515 & -     & Sigmoid Activation  & 1516$\times$1515 \\
8     & Output & 1$\times$1$\times$1515 & -     & Denormalization & 0 \\
Total & -     & -     & -     & -     & 16,271,259 \\
		\hline
\end{tabular}
\smallskip
\caption{Structure of the designed CNN including type, size, option, kernel and trainable parameters of each layer.}
\label{tab:CNN}
\end{table}
\subsection{Direct numerical simulations}
\label{section:numerical}
In order to provide an insight towards the performance of the proposed method compared to the direct numerical simulation approaches, we have used Lattice Boltzmann method (LBM) and pore--scale finite volume solver (PFVS) to calculate the absolute permeability of 3 realistic test samples. Here, we aim to briefly describe the methodology of these numerical methods.
\subsubsection{Pore--scale finite volume solver (PFVS)}
The PFVS method solves an elliptic diffusion equation to obtain the spatial pressure distribution in micro--CT images, hence it estimates the absolute permeability of micro--CT images  \cite{chung2019approximating}. This method is CPU time efficient as it does not require time-stepping to converge to a solution. 
The conservation of mass can be expressed as:
\begin{equation}
\label{eq:FV1}
\nabla . \vec{v}=q
\end{equation}
where $\vec{v}$ is the velocity vector and $q$ represents source and sink terms, which is assumed to be zero. We assign each voxel a local conductivity $w$, calculated as outlined by Chung \textit{et al.} \cite{chung2019approximating} to relate the velocity to the pressure gradient: 
\begin{equation}
\label{eq:FV2}
\vec{v}=-w \nabla P
\end{equation}
where $P$ is pressure. Combining equations \ref{eq:FV1} and \ref{eq:FV2} gives an elliptic equation:
\begin{equation}
- \nabla . (w \nabla P) =0
\end{equation}
This equation is solved with prescribed constant pressure (Dirichlet) boundary conditions on the inlet and outlet by using Two Point Flux Approximation (TPFA) \cite{aarnes2007geometrical,lie2012open} and Finite Volume Methods with an Algebraic Multi-Grid (AMG) solver. All solid voxels are removed from the system of equations, resulting in a smaller system matrix.
Once the spatial pressure distribution and velocity profile (and subsequently, the flow rate) are calculated by equations \ref{eq:FV2}, the absolute permeability is estimated from:
\begin{equation}
K=\frac{N_x}{N_y N_z R}\frac{Q \mu}{\Delta P}
\end{equation}

The length of the system is expressed as the voxels multiplied by the resolution of the image ($R$). Hence, the number of voxels in the main flow direction can be defined as $N_x$ and the other two directions are $N_y$ and $N_z$, $Q$ is flow rate ($m^3/s$), $\mu$ is fluid viscosity ($Pa.s$), $\Delta P$ is pressure difference across the image ($Pa$), imposed as a boundary condition.
Under the assumptions of laminar incompressible flow and no--slip boundary condition, local conductivity ($w$) is defined as a weighting function representing the conductivity of a voxel for fluid flow. Local conductivity is a function of two variables, the largest inscribed radius of the flow channel ($r_{max}$) and the distance from the solid wall ($r$):
\begin{equation}
w= \alpha R^2 \frac{\rho}{8 \mu}(2 d_{max}d-d^2)
\end{equation}
where $w$ is local conductivity, $\alpha$ is the shape factor, $R$ is resolution of image ($m$), $d_{max}$ is digital equivalent of the largest inscribed radius, $d$ is digital equivalent of radial distance from the inner wall, $\rho$ is fluid density ($kg/m^3$), and $\mu$ is fluid viscosity ($Pa.s$).

\subsubsection{Lattice Boltzmann method (LBM)}

In another direct numerical simulation approach, flow within the pore space is calculated by the Lattice Boltzmann Method (LBM) using a Multi-Relaxation Time scheme  \cite{da2019computations,mcclure2014novel,wang2020accelerated} in D3Q19 quadrature space in order to eliminate spurious parameter coupling between viscosity and permeability that occurs with Single Relaxation Time and improve stability in high velocity pore throats. LBM reformulates the Navier-Stokes Equations by numerically estimating the resulting continuum mechanics from underlying kinetic theory. 
The kinetics of a bulk collection of particles within a control volume is estimated with a vector velocity space $\xi_q$ and velocity distributions $f_q$. For each velocity space vector $\xi_q$, the velocity component in the specified direction is given by $f_q$. The momentum transport equation at location $\vec{x_i}$ over a timestep $\delta t$ relies on a collision operation $J$ which recovers the Navier Stokes Equation, and outlined in detail by McClure \textit{et al.}  \cite{mcclure2014novel}.
\begin{equation}
f_q (\vec{x_i}+\vec{\xi_q} \delta t,t+\delta t)=f_q (\vec{x_i},t)+J(\vec{x_i},t)
\end{equation}
Single phase flow is simulated within the pore space of the segmented test samples until steady state conditions are reached. In the present study we continue the simulations until the change in permeability over 1,000 LBM timesteps is less than 1e$ -5 $. All samples are simulated with a constant pressure drop between the inlet and outlet, and wall boundary conditions are imposed along the other sides to avoid geometric inconsistencies associated with periodic boundary conditions.

\section{Results and Discussions}
In this section, three main outcomes of this study are discussed. Initially we describe the significance and applications of the present dataset of the porous material and then we focus on the statistical lessons learned by examining cross--correlations of the dataset features. Finally, the accuracy of the features estimated by the model will be checked on the testing samples to demonstrate the capability of DeePore workflow for rapid characterization of the porous material. 

\subsection{Porous material dataset}
In this research we have generated a comprehensive dataset of semi--real micro--porous structures with 17700 samples and a wide range of morphological, hydraulic, electrical and mechanical features are calculated for each of the samples. The main application of this dataset is to be used as the raw material for more advanced machine learning studies on porous material. In addition to the raw 3-D geometries, Python codes, extracted pore networks and all calculated characteristics are available in the public domain for replication and improvements in future studies \footnote{GitHub Repository: \url{https://github.com/ArashRabbani/DeePore}}.  

\subsection{Statistical lessons learned}

\begin{figure}[H]
	\centering\includegraphics[page=\value{fig_num},width=1\linewidth]{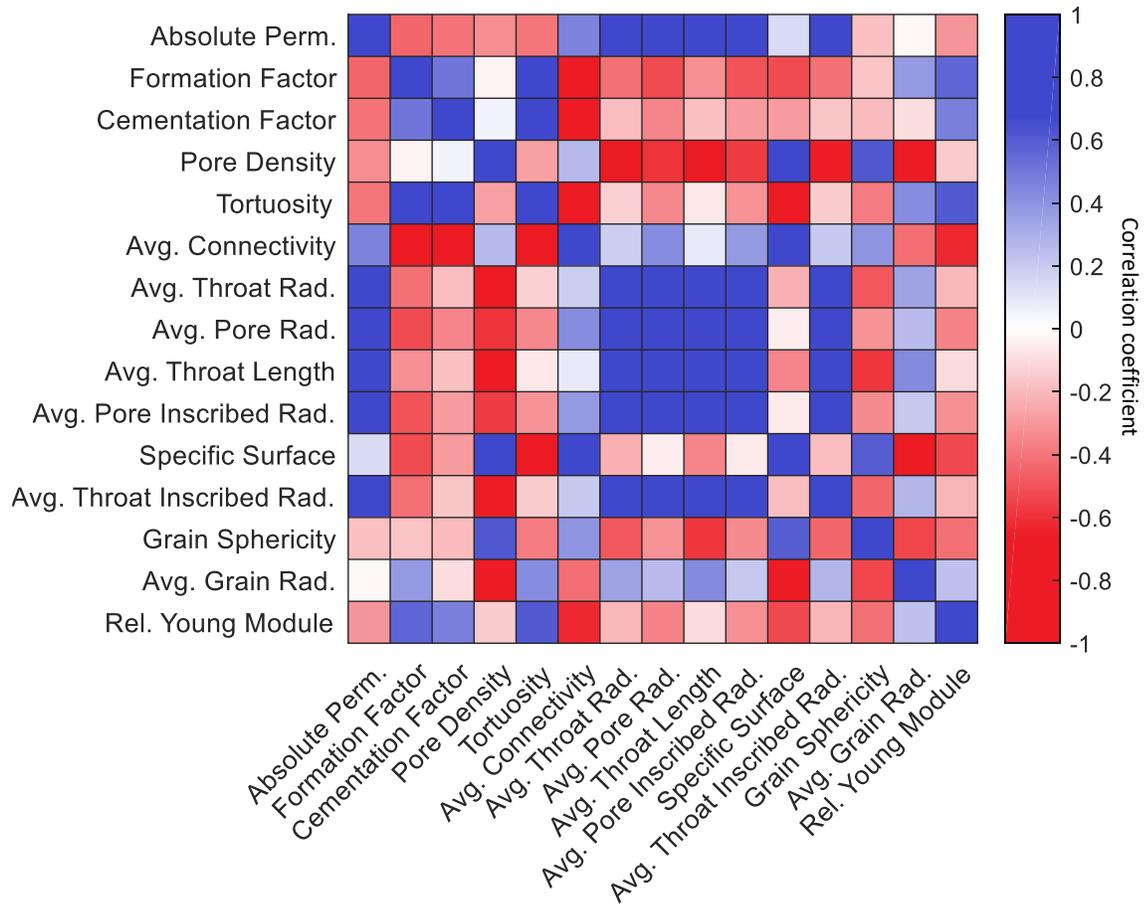} \stepcounter{fig_num}
	\caption{Heatmap of the cross--correlations between the physical properties of the porous material, blue color indicates that two variables are highly correlated, white color indicates that variables are not statistically related and red color denotes a strong inverse correlation.}
	\label{fig:heatmap}
\end{figure}

Considering the large number of analyzed samples of porous material, we have created a rich dataset to investigate the existing trends and relationships among the calculated features. Binary correlation coefficients of 15 single-value features are visualized in Fig. \ref{fig:heatmap} as a heat map to summarize the statistical significance of cross-parameter relationships. Pure blue color at the intersection of two parameters indicates strong correlation and pure red color shows a strong inverse correlation between them. As an example, absolute permeability of porous media is well correlated with average pore--throat radius which is expected based on the available literature \cite{rezaee2006relationships}, while it does not have a significant relationship with average grain radius and finally it has an inverse relationship with relative Young module of the porous material. This relationship is physically justifiable considering that large values of relative Young module indicate a tight and consolidated structure of porous material \cite{yeheskel2000new} which leads to lower permeability. Although, many of these relationships have been widely investigated in the literature \cite{mortensen1998relation,timur1968investigation,pittman1992relationship}, having a diverse range of them in a single map (Fig. \ref{fig:heatmap}), could provide a concise but broad insight about porous material characteristics.  

\subsection{Data distribution}
\begin{figure}[H]
	\centering\includegraphics[page=\value{fig_num},width=1\linewidth]{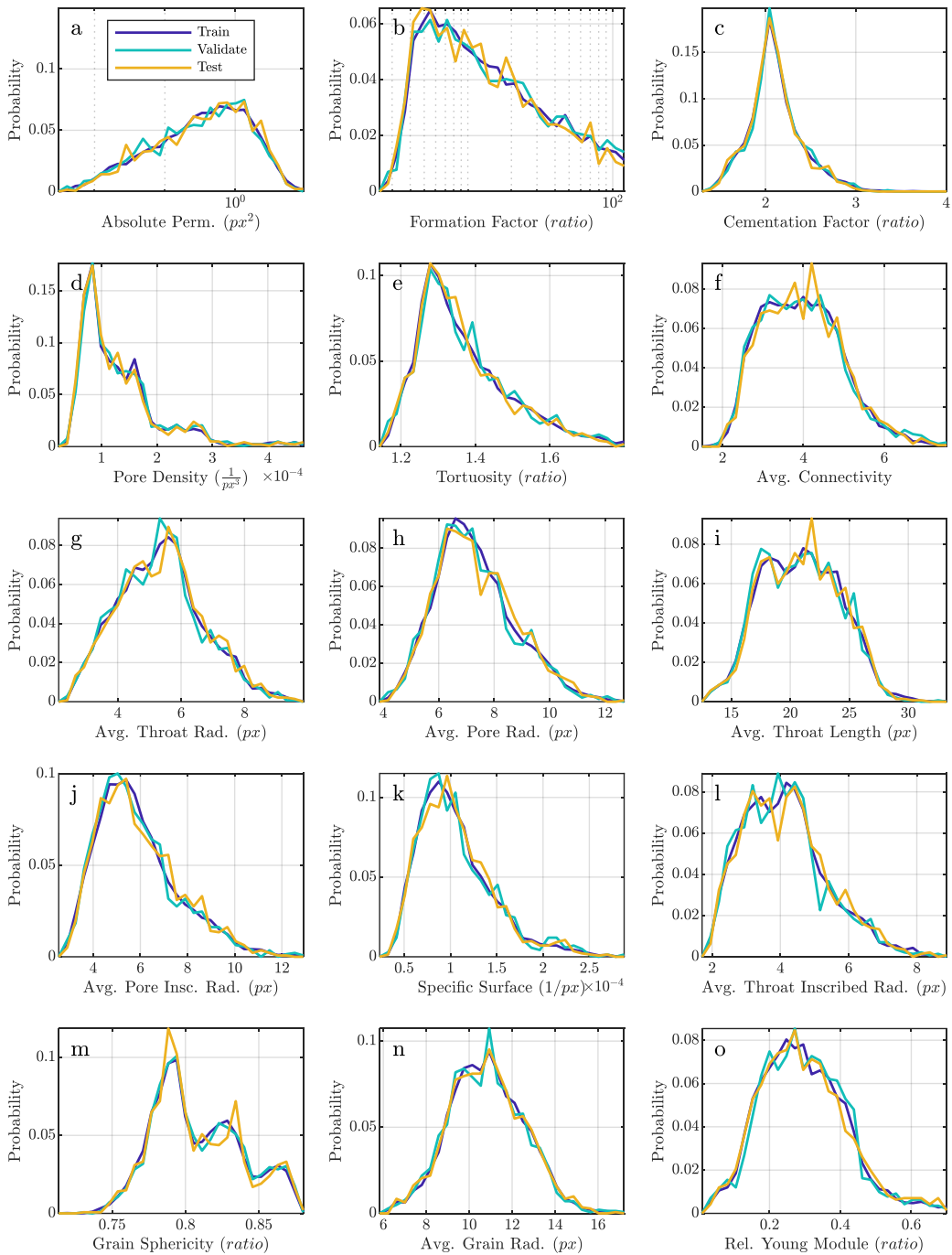} \stepcounter{fig_num}
	\caption{Relative frequency distribution of 15 single--value features for each bundle of the training, validation and testing data.}
	\label{fig:data_stat}
\end{figure}

The data is randomly split by shuffling into three sections with the proportions of 80 \%, 10 \%, and 10 \% respectively for training, validation, and testing processes. Relative frequency distribution of the training, validation, and test data are plotted in Fig. \ref{fig:data_stat} to demonstrate similarity/difference of the data statistics. We have run two--sample Kolmogorov--Smirnov (K--S) test to check if the distributions of training and validation/test data are statistically similar. K--S distances of the tests are provided in Appendix B and if the value is closer to 1, it indicates that the two compared distributions are dissimilar. Based on the obtained results, 97 \% of the K--S distances are below 0.03 which shows that data shuffling and sample selection are unbiased which is favourable in terms of the training robustness. Fig. \ref{fig:data_stat} illustrates the three distributions of training, validation and testing datasets for 15 single--value features modelled in this study.

\subsection{Alternative models}
\begin{figure}[H]
	\centering\includegraphics[page=\value{fig_num},width=.5\linewidth]{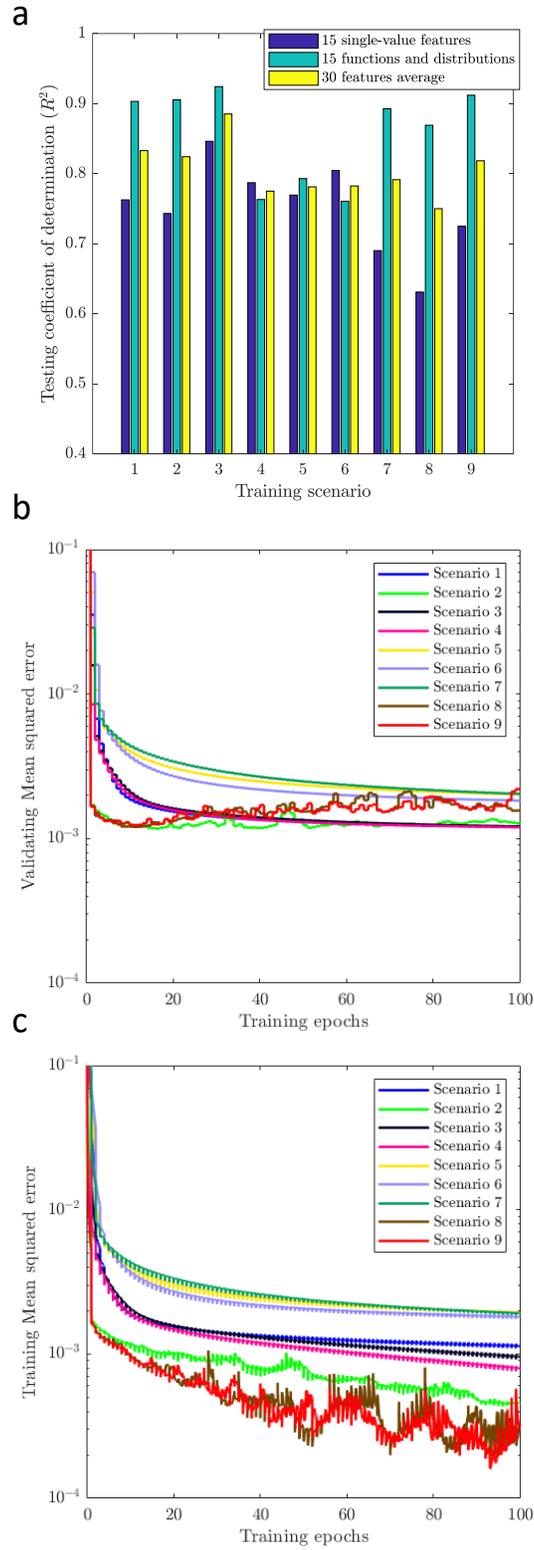} \stepcounter{fig_num}
	\caption{Comparing 9 training scenarios in terms of testing coefficients of determination (r-squared) (a), validating mean squared error per epoch (b), and training mean squared error per epoch (c).}
	\label{fig:loss}
\end{figure}

In order to fine--tune the model structure and select the best options, we have tested 9 training scenarios of which they were more likely to fit the purpose. Different number of filters in each convolution layer, corresponding kernel sizes, optimizers, and loss functions used for each scenario are presented in table \ref{tab:compare_scenarios}. Additionally, the total averaged $R2$ of the models based on the testing dataset is calculated to be used as evaluation criterion. Also, the averaged $R2$ is calculated for both group of single--value features and functions illustrated in Fig. \ref{fig:loss}--a. It has been found that the accuracy of the models is dependant to both loss function and network structure. Four different loss functions have been tested through 9 scenarios:
\begin{itemize}
	\item \textbf{Mean squared error (MSE)}: this is the simplest version of loss function and by using it we assume that all the 1515 elements of the CNN output array share the same weight in the training process.
	\begin{equation}
	Loss=\frac{1}{1515}\sum_{i=1}^{1515} (y{_i}-y'{_i})^2
	\end{equation}
	where, $y_i$ is the actual value and $y'_i$ is the predicted value of the $i_{th}$ elements of the output array.
	\item \textbf{Weighted MSE}: this type of loss function indicates that the 15 single--value outputs should have same weight as the 15 functions and distributions each of which with 100 elements. Thus, MSE is calculated for each of the function elements then it is averaged over the 100 elements and summed up with the 15 single--value errors as follows: 
	\begin{equation}
	Loss=\frac{1}{15}\sum_{i=1}^{15} (y{_i}-y'{_i})^2+\frac{1}{15}\sum_{i=16}^{30} \frac{1}{100}\sum_{j=1}^{100}(y_{ij}-y'_{ij})^2
	\end{equation}
	\item \textbf{Binary cross entropy}: This loss function is regularly used for evaluating the accuracy of a binary classification, while it has been found in this paper to be effective for the functions and numbers that revolve around the values between 0 and 1. This loss function performs a maximum likelihood estimation based on the Kullback–Leibler divergence \cite{rubinstein1999cross} and it performs better than MSE in the cases with large difference in the order of magnitude due to the logistic formulation as follows: 
	\begin{equation}
	Loss=\frac{1}{1515}\sum_{i=1}^{1515} y{_i}(-\log(y'{_i}))+(1-y{_i})(-\log(1-y'{_i}))
	\end{equation}
	\item \textbf{Weighted binary cross entropy}: this loss function behaves similar to the regular binary cross entropy, but it is normalized over 100 elements of the 15 functions and distributions to avoid unnecessary additional influence of them in model training due to the higher quantity compared to the single--value outputs. The formulation is presented as follows:
	\begin{multline}
	Loss=\frac{1}{15}\sum_{i=1}^{15} y{_i}(-\log(y'{_i}))+(1-y{_i})(-\log(1-y'{_i}))	\\+\frac{1}{15}\sum_{i=16}^{30}\frac{1}{100}\sum_{j=1}^{100}y_{ij}(-\log(y'_{ij}))+(1-y_{ij})(-\log(1-y'_{ij}))
	\end{multline}
\end{itemize}     

In addition to the loss functions discussed above, two types of optimizers have been utilized in the alternative scenarios namely Adam and RMSprop. These are two adaptive gradient-based optimization method for stochastic objective functions \cite{kingma2014adam,zou2019sufficient}. These methods store an exponentially decaying average of past squared gradients to utilize it for future estimations. The advantage of Adam is the fact that it also keeps the momentum of past gradients which helps it for better estimation of higher order behaviours \cite{ruder2016overview}. We have used learning rates of $10^{-5}$ and $10^{-3}$ for RMSprop and Adam optimizers, respectively.

Based on the model performances presented in Table \ref{tab:compare_scenarios} and Fig. \ref{fig:loss}--a, it can be concluded that a simple mean squared error is suitable to be used as the loss function for the present dataset. Also, the comparison shows that weighted MSE underperforms the simple MSE in terms of coefficient of determination.
($R^2=0.885$ among the alternative models). In addition to the r--squared comparison, we have illustrated the MSE of the validation and training datasets at the end of each epoch (Fig. \ref{fig:loss}--b and c). As it can be seen, training and validation curve of Scenarios 3 are stable when approaching to 100 epochs. In addition, they are converging almost to a same MSE while in Scenarios 2, 8, and 9 overfitting are observed. Considering all discussed criteria, Scenario 3 is recommended and we use it as the predictor model for results presented hereafter.    

Another alternative approach to perform this modeling is the use of separated CNNs for each of the 30 output features. Theoretically, this approach can lead to a better performance while for the purpose of this study, it is not without flaws. Use of 30 different CNNs not only increases the storage size and computational burden of the model up to several times, but also it offers an unnecessary level of accuracy which is beyond the accuracy of the dataset. Considering the fact that the PNM--obtained ground truth data such as the one for permeability contains around 5 to 30 \% of simplification error \cite{baychev2019reliability}, the present commingled structure of CNNs is sufficiently accurate. In addition, the present light implementation of the model makes the online and client--side predictions more viable. 

\begin{table}[H]
	\footnotesize
	\centering
	\begin{tabular}{c| c| c| c| c | c}
		\hline
Scenarios & Number of filters & Kernel sizes & Optimizer & Loss function  
                       & Averaged $R^2$ \\
		\hline
1         & 6,12,18           & $8^2$,$4^2$,$2^2$        & RMSprop   & MSE                          & 0.832          \\
2         & 6,12,18,24        & $8^2$,$4^2$,$2^2$,$2^2$      & RMSprop   & MSE                          & 0.824          \\
3         & 12,24,36          & $3^2$,$3^2$,$3^2$        & RMSprop   & MSE                          & 0.885         \\
4         & 6,12,18           & $8^2$,$4^2$,$2^2$        & RMSprop   & Weighted MSE                 & 0.775          \\
5         & 6,12,18,24        & $8^2$,$4^2$,$2^2$,$2^2$      & RMSprop   & Weighted MSE                 & 0.781          \\
6         & 12,24,36          & $3^2$,$3^2$,$3^2$         & RMSprop   & Weighted MSE                 & 0.782          \\
7         & 6,12,18           & $8^2$,$4^2$,$2^2$        & Adam      & Binary cross entropy          & 0.791          \\
8         & 6,12,18           & $3^2$,$3^2$ ,$3^2$        & Adam     & Binary cross entropy          & 0.749          \\
9         & 6,12,18           & $3^2$,$3^2$ ,$3^2$       & Adam     & Weighted binary Cross entropy & 0.818      \\   
	\hline
	\end{tabular}
\smallskip
\smallskip
\caption{Comparing 9 different training scenarios in terms of filter number and kernel sizes, optimizer, loss and coefficient of determination ($R^2$). Scenarios 3 show the best performance and have been selected to be used as the DeePore ANN structure.}
\label{tab:compare_scenarios}
\end{table}

\subsection{Model performance}
Model training is performed both in CPU and GPU architectures. CPU--based computations are done on a machine with four 3.2 GHz. Intel Xenon processors and 32 GB of memory. This arrangement enables the model to be trained in 3 to 4 minutes per epoch. In addition, prediction of the porous media features based on the trained model takes 9.23 ms per sample on average which is 4 to 6 orders of magnitude faster if the features are to be extracted by physical simulations whether PNM or direct numerical simulation methods \cite{rabbani2019hybrid}.

Using GPU to accelerate the training and prediction stages can make this workflow even faster. Using an Nvidia GeForce GTX 1660 Ti with Max-Q Design Graphic Card with Compute Capability of 7.5 and 6 GB of memory, we are able to train the model around 4 times  faster, in which each epoch would take around 45 s to finish. Also prediction of the test dataset can be performed with the speed of 0.379 ms per sample on average which is 24 times faster than the CPU--based instance and it is a considerable improvement. 

At each epoch of training the validation and training losses are measured in terms of mean squared error to ensure that over--fitting is not occurring and training has reached an optimum point (Fig. \ref{fig:loss}--b and c). In our case, increasing the number of epochs to more than 100, it hardly reduces the validation loss less than $10^{-3}$ in a stable manner. So, we have stopped the training at 100 epochs as visualized in Fig. \ref{fig:loss}--b and c.

Using the trained CNN model we have estimated a wide range of porous material characteristics on the 10\% of the data which are not used in the training or validation processes. The average determination coefficients ($R^2$) of the reference and estimated features for the test data are 0.846 and 0.924, respectively, for single--value features (rows number 1 to 15 of Table \ref{tab:outputs}), and function/distribution features (rows number 16 to 30 of Table \ref{tab:outputs}). Also, the average determination coefficient of all 30 features is 0.885. This overall level of accuracy is reasonably good considering the wide range of porous structures and variety of the predicted features. 

\begin{figure}[H]
	\centering\includegraphics[page=\value{fig_num},width=1\linewidth]{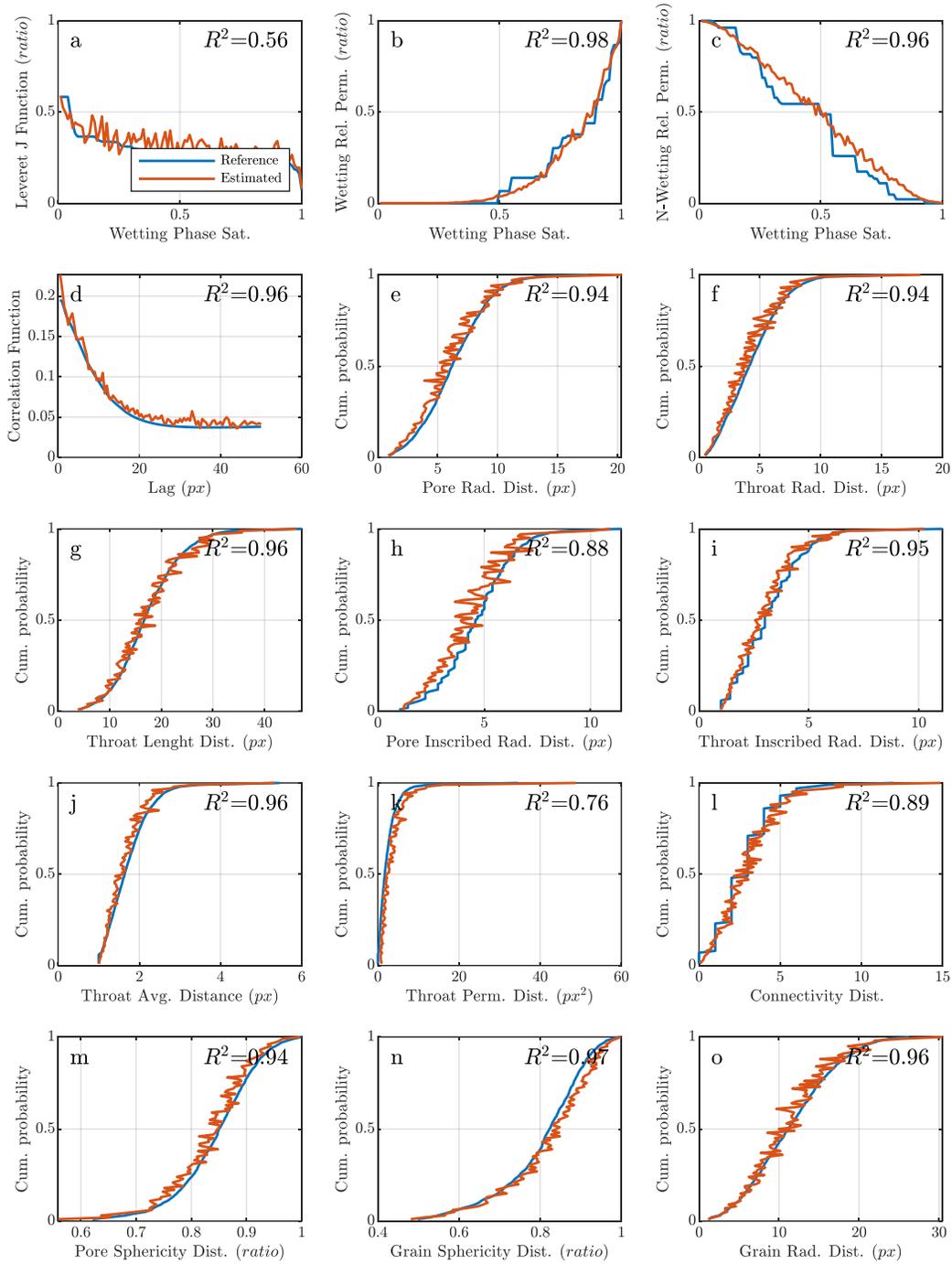} \stepcounter{fig_num}
	\caption{Example comparison of the reference and estimated values of 15 functions and distribution for one of the image samples in the dataset (sample \#1). The average correlation coefficient of the reference and estimated curves is 0.9616.}
	\label{fig:sample_output}
\end{figure}

In order to provide an insight into the sample outputs of the model, we have presented Table \ref{tab:sample_output} and Fig. \ref{fig:sample_output} for discussing single--value features and functions, respectively. In Table \ref{tab:sample_output}, reference values and estimated values are matching with a low level of average relative error at only 5.4\%. If the real spatial resolution of this sample image is assumed to be 5 microns per voxel, consequently, in order to scale the values with the unit of length ($px$) we simply multiply them by $5$. Similarly for the area ($px^2$), values are multiplied by $5^2$. This approach gives us the flexibility to predict porous material characteristics in a wide range of spatial resolutions without changing or re--training the model. A similar approach is used for functions and distribution curves which are depicted in Fig. \ref{fig:sample_output}. The average determination coefficient of all 15 curves shown in this figure for sample \#1 is 0.9417. 

\begin{figure}[H]
	\centering\includegraphics[page=\value{fig_num},width=1\linewidth]{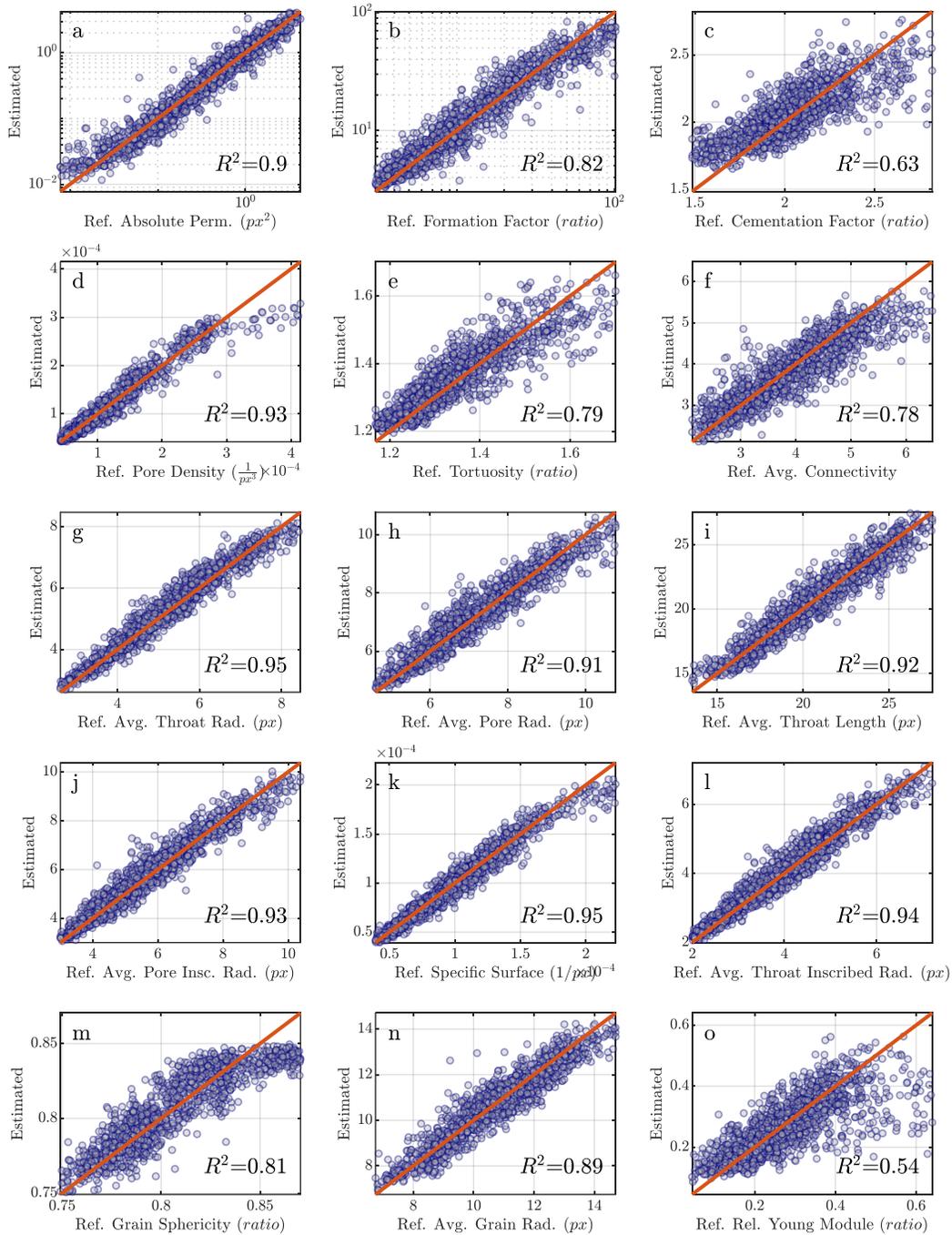} \stepcounter{fig_num}
	\caption{Comparison between the reference and estimated values for 15 single--value features and their correlation coefficients, a to o charts correspond to the features inscribed in Table \ref{tab:outputs} rows from 1 to 15.}
	\label{fig:single_prediction}
\end{figure}

Additionally we have compared the distribution probability of all single--value features to check the correspondence between the reference and estimated data (Fig. \ref{fig:distributions}). This data has been visualized for 1418 testing samples and in the most of the cases both two distributions follow the same pattern, while for parameters like relative Young module and , some deviations are observed (Fig. \ref{fig:distributions}-o). Also, in grain sphericity reference curve, a trimodal behavior is observed while the estimated distribution has failed to match that trend accurately (Fig. \ref{fig:distributions}-m). This distributed data are also visualized in the form of scattered plots in Fig. \ref{fig:single_prediction}. These charts carry the same $R^2$ values as the distribution curves in Fig. \ref{fig:distributions}, while unit--slope lines in Fig. \ref{fig:single_prediction} could give better insight regarding the over or under--estimation of the parameters. We have observed a degree of underestimation for larger values especially in tortuosity, grain sphericity and average connectivity data (Fig. \ref{fig:single_prediction}-e and f). This is probably due to the skewness of the distribution of these parameters in which a long but thin tail toward the larger values do not lead the model to sacrifice its accuracy on the middle--range data for better coverage on the whole domain. In other words, during the training process, model prefers to have higher accuracy for the majority of the data points instead of less accuracy but better coverage on the whole set of points.

\begin{figure}[H]
	\centering\includegraphics[page=\value{fig_num},width=1\linewidth]{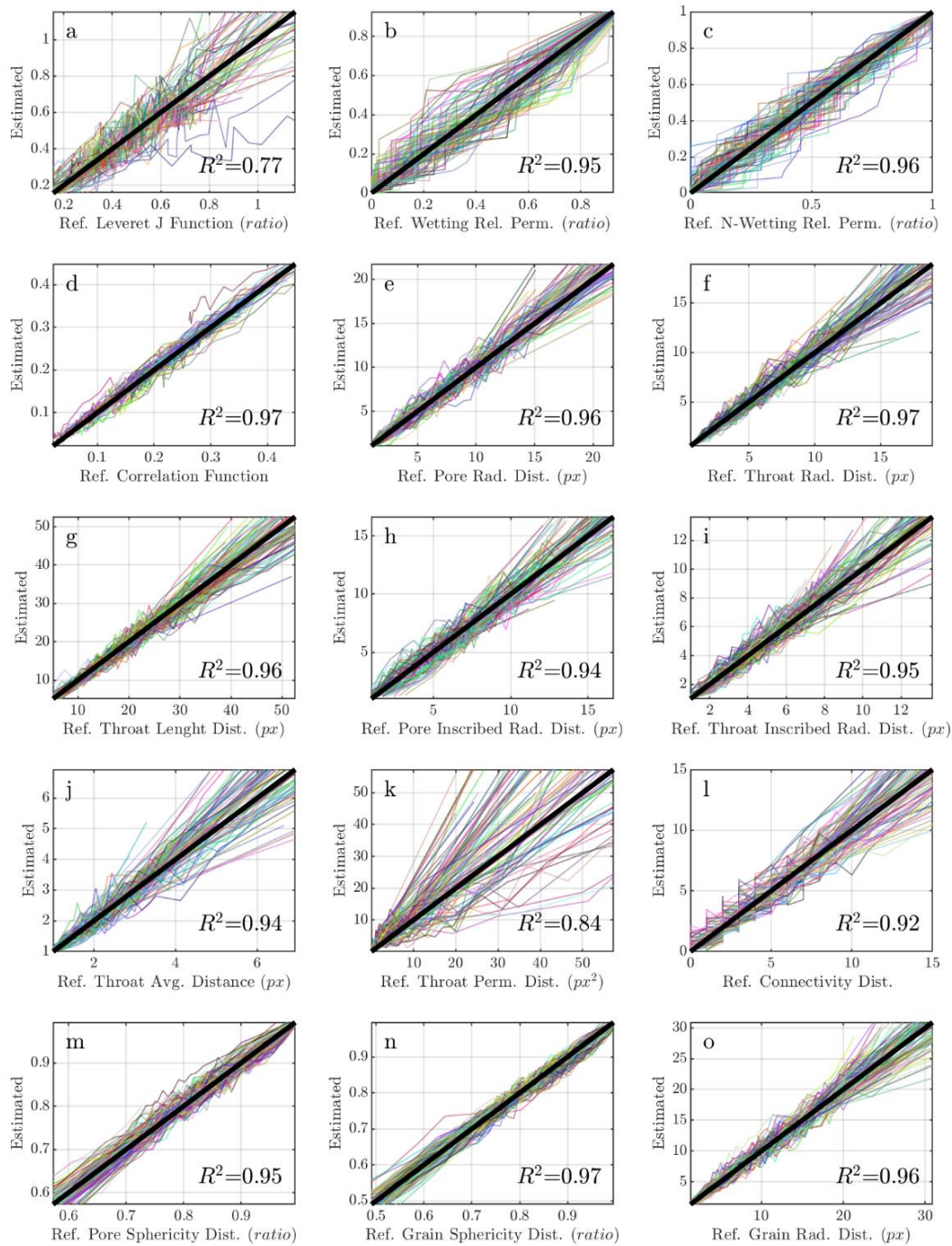} \stepcounter{fig_num}
	\caption{Comparison of the reference and estimated curves for 15 functions and distributions that correspond to the rows of 16 to 30 in Table \ref{tab:outputs} (plotted for randomly selected 100 samples and labeled with random colors).}
	\label{fig:function_prediction}
\end{figure}

Finally, we have plotted the predicted functions and distributions versus reference values for 100 randomly selected samples from the testing pool of data (Fig. \ref{fig:function_prediction}). Also, we have calculated $R^2$ coefficients for each of the plotted lines and the average values are presented in Fig. \ref{fig:function_prediction}. We have found that the data which are directly calculated from the geometrical features of the porous samples such as pore radius distribution (Fig. \ref{fig:function_prediction}-e) are easier to estimate compared to more complicated functions, such as throat permeability distribution (Fig. \ref{fig:function_prediction}--k) that are obtained from a higher level simulation. 

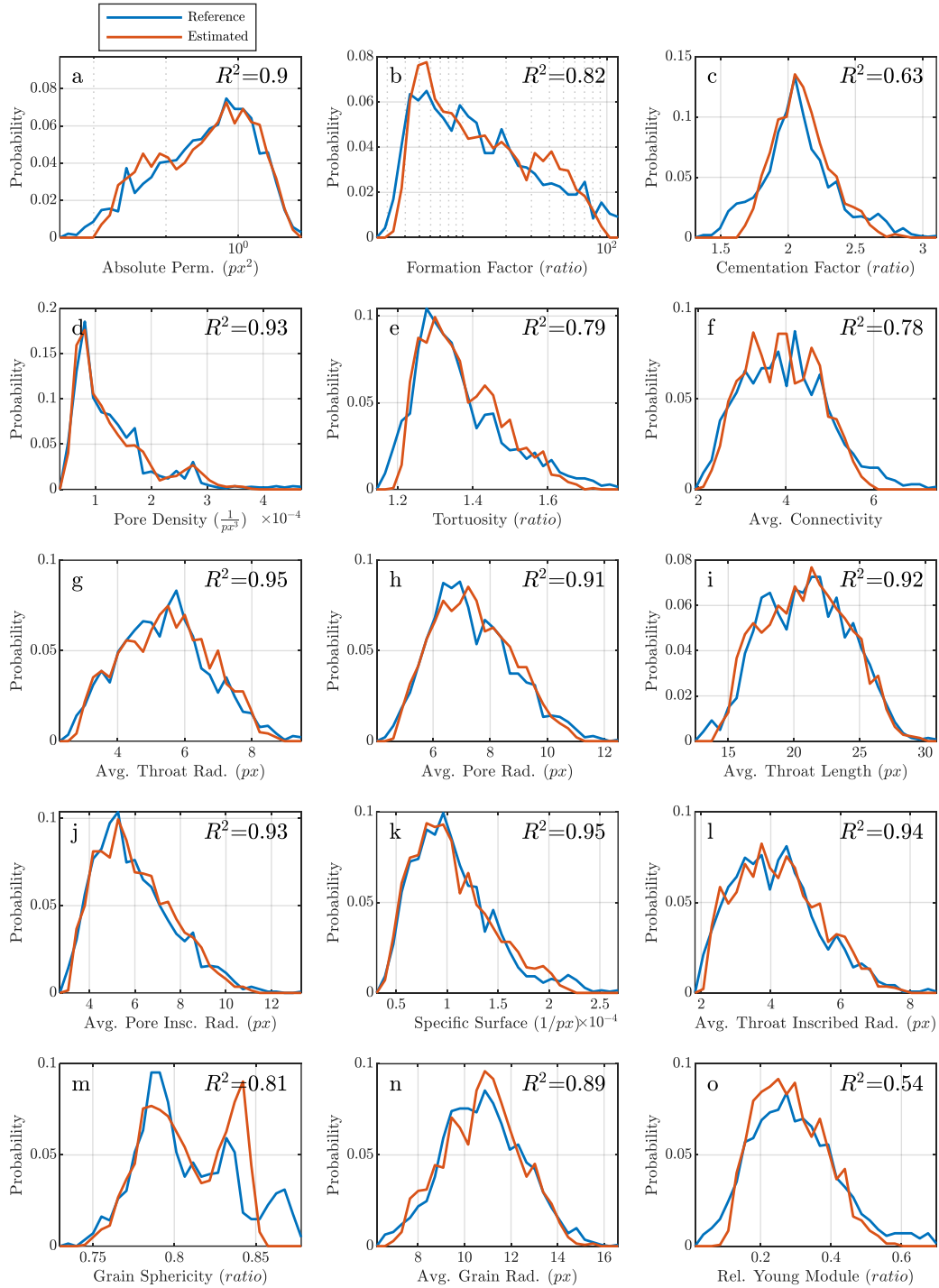
\begin{table}[H]
	\footnotesize
	\centering
	\begin{tabular}{c| c| c| c| c}
		\hline
		Num. &  Feature & Reference Value & Estimated Value & Relative Error (\%) \\
		\hline
1  & Absolute permeability ($px^2$)          & 0.544    & 0.573    & 5.39  \\
2  & Formation factor (ratio)             & 6.541    & 5.542    & 15.28 \\
3  & Cementation factor (ratio)           & 2.000    & 1.980    & 1.00  \\
4  & pore density ($1/px^3$)                 & 1.48E-04 & 1.43E-04 & 3.25  \\
5  & Tortuosity (ratio)                   & 1.244    & 1.264    & 1.62  \\
6  & Average coordination number           & 4.566    & 4.584    & 0.40  \\
7  & Average throat radius (px)           & 4.797    & 5.008    & 4.40  \\
8  & Average pore radius (px)             & 6.952    & 6.965    & 0.18  \\
9  & Average throat length (px)           & 19.722   & 19.602   & 0.61  \\
10 & Average pore inscribed radius (px)   & 5.379    & 5.722    & 6.37  \\
11 & Specific surface (1/px)              & 1.39E-04 & 1.36E-04 & 1.98  \\
12 & Average throat inscribed radius (px) & 3.585    & 3.836    & 7.01  \\
13 & Grain sphericity (ratio)             & 0.807    & 0.819    & 1.49  \\
14 & Average grain radius (ratio)         & 8.575    & 9.039    & 5.41  \\
15 & Relative young module (ratio)        & 0.219    & 0.160    & 26.66 \\

		\hline
		& Average & - & - & 5.4 \\  
		\hline
	\end{tabular}
	\smallskip
	\caption{Example comparison of the reference and estimated values of 15 single--value features for one of the image samples in the dataset (test sample \#1).}
	\label{tab:sample_output}
\end{table}

\begin{figure}[H]
	\centering\includegraphics[page=\value{fig_num},width=1\linewidth]{Figure_cropped.pdf} \stepcounter{fig_num}
	\caption{Comparison between the reference and estimated distributions of the 15 single--value features, a to o charts correspond to the features inscribed in Table \ref{tab:outputs} rows from 1 to 15.}
	\label{fig:distributions}
\end{figure}

\subsection{Independent model verification}
\begin{figure}[H]
	\centering\includegraphics[page=\value{fig_num},width=1\linewidth]{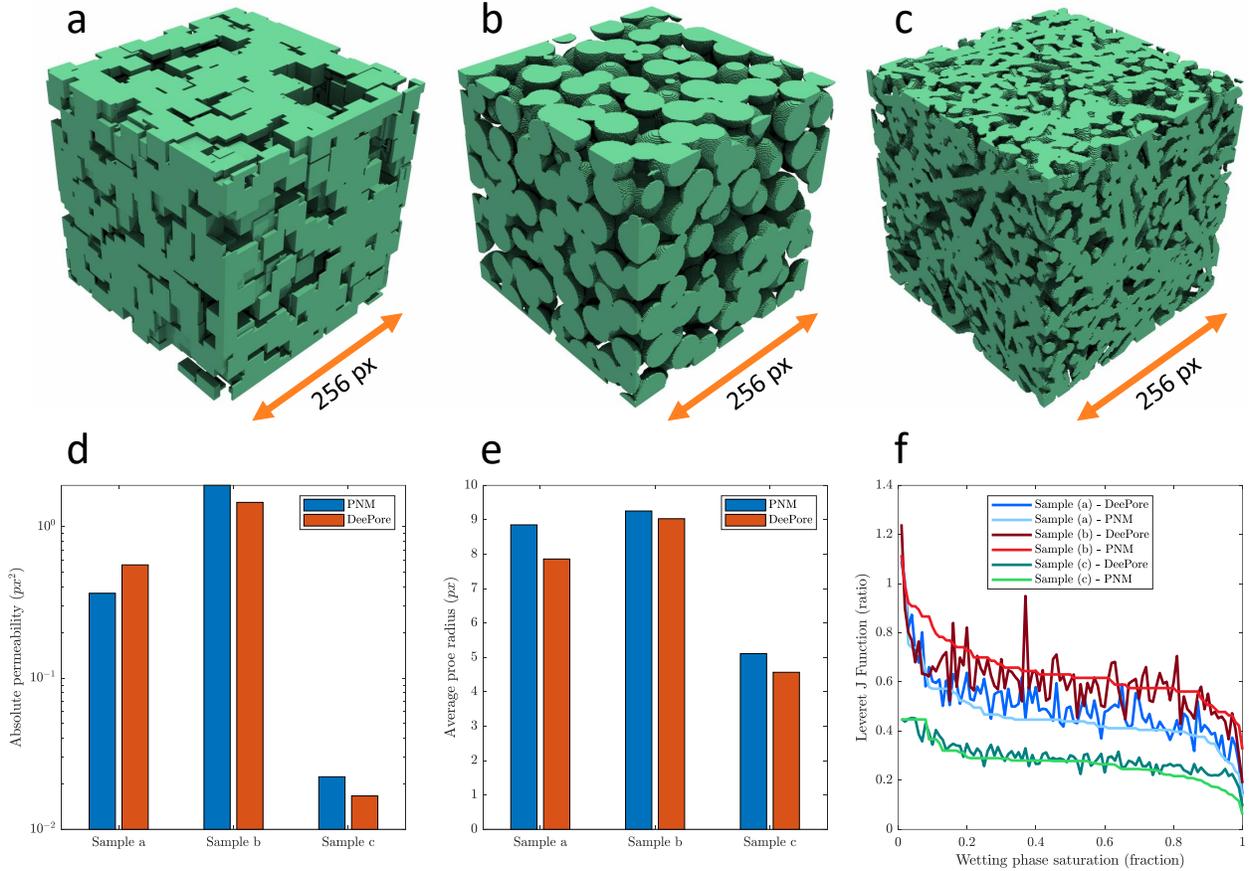} \stepcounter{fig_num}
	\caption{Independent verification of model using three artificial porous structures completely outside the training, validation and testing dataset. (a) overlapped pack of cubes, (b) partially overlapped pack of spheres, (c) fibrous media of straight cylindrical rods, (d) comparing the absolute permeability of three images obtained by PNM and DeePore, (e) comparing the average pore radius of three images obtained by PNM and DeePore, and (f) comparison of Leveret J function of three porous samples simulated by PNM and estimated by DeePore.}
	\label{fig:verify}
\end{figure}

In order to ensure that the model is not over--trained with the augmented dataset of semi--realistic images, we have created three fundamentally different and independent images with no similar structure inside the dataset to check if the model has implicitly learned the physics instead of only memorizing different textures and corresponding features. The results are almost as good as the testing dataset used to check the model performance in the previous subsection in terms of r--squared. By lumping 15 single--value features of 3 verification images, r--squared of 0.916 is obtained which shows that the DeePore predictions are noticeably similar to the pore network modeling simulations. In Fig. \ref{fig:verify} we have visualized three porous samples: (a) the medium is made up of overlapping cubes with no offset limitation, (b) a packing of spheres with overlapping length limited by the half of the spheres' radii, and (c) a fibrous medium made by straight cylindrical rods with 10 voxels radius. To provide some examples of the estimated features, we have compared the absolute permeability, average pore radius and Leverett J function of three constructed samples simulated by PNM and estimated by DeePore (Fig. \ref{fig:verify}--d to f, respectively). Averaged relative error of the permeabilities obtained is around 34 \% which is reasonable considering the wide range of variation of this variable. Also, for average pore size this error is around 8.1\% that is not out of expectation due to the high predictability of pore size from images. Finally, Leverett J function of three samples has been reasonably accurately estimated by DeePore, although a high level of noise is observed in the prediction. The shape of the fluctuations in Fig. \ref{fig:verify}--f are visually similar to the white unbiased noise which can be easily cancelled by performing a moving average or Gaussian filter. 

\subsection{Verification with experimental data}
\begin{figure}[H]
	\centering\includegraphics[page=\value{fig_num},width=1\linewidth]{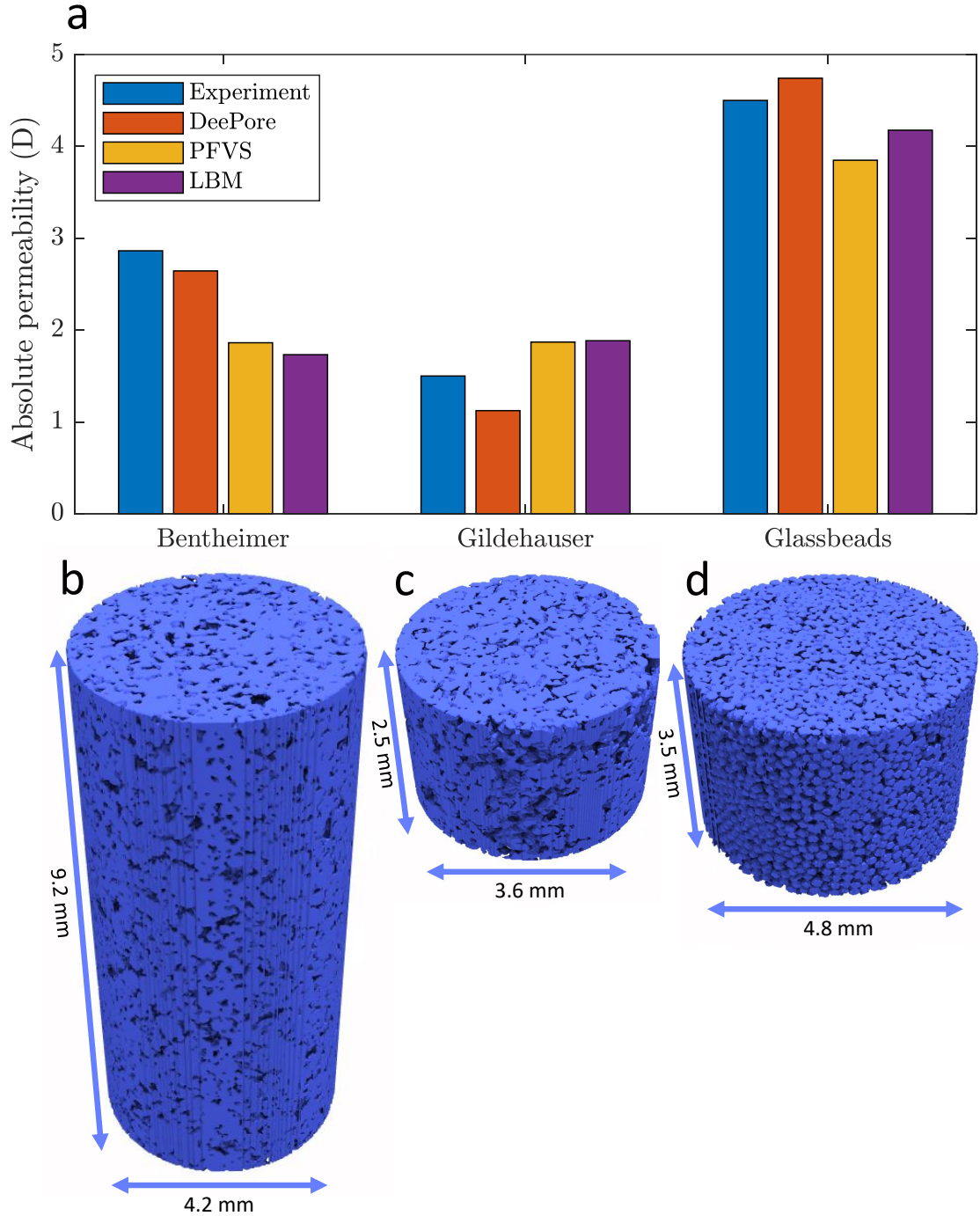} \stepcounter{fig_num}
	\caption{Experimental verification of the proposed model by comparing absolute permeabilities with experimental values as well as direct numerical simulation, (a) Absolute permeabilities of Bentheimer sandstone, Glidehauser sandstone, and glass beads obtained from 4 different sources:  experiment, DeePore, PFVS, and LBM. (b) Segmented image of the tested Bentheimer sandstone with resolution of 7 $\mu m/voxel$. (c) Segmented image of the tested Glidehauser sandstone with resolution of 4.4 $\mu m/voxel$. (d) Segmented image of the tested glass beads with resolution of 6.5 $\mu m/voxel$.}
	\label{fig:exp_verify}
\end{figure}

Considering the fact that the dataset of images used in this study have been virtually augmented, a verification with realistic images can provide a better insight towards the applicability of the proposed method for real world problems. In this regard, three tomography images with available pore--scale experiments of absolute permeability have been used. Additionally, we have employed two direct numerical simulation methods of PFVS and LBM (described briefly in Section~\ref{section:numerical}) to compare the performance of the proposed method. Absolute permeabilities obtained from experiment, Deepore, PFVS and LBM are illustrated in Figure \ref{fig:exp_verify}. Also in this figure we show the segmented images of three samples used namely Bentheimer sandstone \cite{ramstad2012relative}, Glidehauser sandstone \cite{rucker2015connected}, and glass beads \cite{hasan2020direct}. In order to find the permeability of the samples in the cylinder's axial direction, we divide the image into 2--D slices perpendicular to the cylinder main axis. Then using a $256^2$--voxels 2-D sliding window we take subsamples to use as DeePore feed. The overal permeability of a slice is determined by arithmetic averaging over all subsamples and this process repeats for all slices of the 3-D image. Next, minimal slice permeability over the whole length of the sample is reported as the sample directional permeability considering the fact that flow capacity is mainly controlled by the tightest openings and bottlenecks. By comparing the three predicted absolute permeabilities with the experimental values it can be concluded that DeePore is predicting the permeability in a good agreement with the direct numerical simulation methods which are considerably more computationally expensive. The average relative error of tested methods compared to the experiment are 13 \%, 25 \%, and 24\%, respectively for DeePore, PFVS, and LBM. Although there could be some sources of uncertainty (such as images' improper segmentation, or experimental and imaging errors), the general conclusion is that based on the tested realistic tomography images, DeePore absolute permeabilities are in good agreement with experimental measurements.  
 
\section{Conclusions and Future Studies}
\label{sec:con}
In order to deliver the present DeePore package, we have generated a dataset of semi--real 3--D images of porous structures with 17700 samples and 30 features which are physical characteristics of porous material. Using a regression CNN coupled with two dense layers, a fast and comprehensive characterization of 3--D porous material images is implemented. In summary, the following conclusions can be drawn based on the findings of this research:

\begin{itemize}
	\item  A physically diverse dataset of micro--porous structures are generated based on texture transformation and porosity manipulation of 60 original tomography images and a wide range of morphological, hydraulic, electrical and mechanical features are extracted for each sample. The dataset is publicly available and can be used for future studies.
	\item A dimensionless approach has been presented to extract or predict the porous material features without being affected by the spatial resolution of the images. 
	\item  A statistically accurate feature estimation is implemented using a designed feed--forward CNN by training the model for 100 epochs while over--fitting is avoided. 
	\item A cross--correlation heat map of the 15 porous material features are presented that gives a concise but broad insight into the significance of relationships between each pair of the extracted features. 
	\item  The average coefficient of determination ($R^2$) for all 30 extracted features is 0.885 which is significant considering the diverse range of porous morphologies and features.
	\item  Via a GPU--based architecture we are able to predict the above--mentioned features for each $256^3$ voxels binary image of porous material in 0.379 ms on average. This high speed of prediction enables us to tackle larger image sizes in future studies. 
	\item To provide an independent model verification, three images outside of the datasets are generated with textures almost unseen in the augmented dataset. The model predictions were almost as good as the testing dataset used to check the model performance in terms of r--squared. This observation indicates that the dataset is diversified enough to avoid false accuracy due to the similarity of the training and test results. Also, considering the novel texture of the verification samples, one can conclude that the model is implicitly learning the physics of the estimated features rather than blindly memorizing the textures. 
	\item In order to provide experimental verification for the proposed method, absolute permeabilities of 3 realistic porous samples have been compared with the measured values in the laboratory and DeePore predictions had around 13\% relative error with is noticeable comparing to the accuracy of the direct numerical simulation methods such as PFVS and LBM.   
	
\end{itemize}

It is noteworthy that each of the physical simulation methods are subject to uncertainty due to some inherent simplification assumptions in pore network modeling. Consequently, it is recommended to use more accurate simulation methods such as direct numerical models to generate the ground truth dataset based on the current set of 3--D images in the future studies.

\section*{Acknowledgement}
The authors thank the University of Manchester for the President’s Doctoral Scholarship Award 2018 awarded to Arash Rabbani to carry out this research. Also, special thanks to Prof. Vahid Niasar for sharing the glass beads image that has been used in the present study.

\section*{Author Contribution Statement}
Arash Rabbani has developed the idea of this research, executed the research by computer code development and written the main body of the article. Reza Shams contributed to execution of research through trouble shooting of the computer codes. Ying Da Wang executed the LBM simulations and contributed in writing the article. Traiwit Chung executed the PFVS simulations and contributed in writing the article. Masoud Babaei supervised the research and contributed in writing the article.
\newpage
\section*{Appendix A}
Here we have listed the real tomography images of porous material used to generate the dataset using the data augmentation method. The size and spatial resolutions mentioned in this table are modified by resizing/cropping the original images to fit the purpose of this study.

\scriptsize

\begin{longtable}[c]{c| c |c |c |c |c}
	\hline
	Num.     & Name & Resolution & Modified  & Type &  DOI \\
          &    & ($\mu m/px$) &  size ($px^3$) &  &   \\
\hline

		1     & Berea \#1 & 11.72 & $256^{3}$ & Sandstone & https://doi.org/10.1016/j.advwatres.2012.03.003 \\
		2     & Berea \#1 & 8.35  & $256^{3}$ & Sandstone & http://dx.doi.org/10.1103/PhysRevE.80.036307 \\
		3     & Berea \#2 & 3.25  & $256^{3}$ & Sandstone & https://doi.org/10.1016/j.advwatres.2012.03.003 \\
		4     & Berea \#3 & 3.25  & $256^{3}$ & Sandstone & https://doi.org/10.1016/j.advwatres.2012.03.003 \\
		5     & Berea \#4 & 3.25  & $256^{3}$ & Sandstone & https://doi.org/10.1016/j.advwatres.2012.03.003 \\
		6     & Berea \#5 & 4.33  & $256^{3}$ & Sandstone & https://doi.org/10.1016/j.advwatres.2012.03.003 \\
		7     & Berea \#6 & 4.33  & $256^{3}$ & Sandstone & https://doi.org/10.1016/j.advwatres.2012.03.003 \\
		8     & Berea \#7 & 4.33  & $256^{3}$ & Sandstone & https://doi.org/10.1016/j.advwatres.2012.03.003 \\
		9     & C1    & 4.45  & $256^{3}$ & Carbonate & http://dx.doi.org/10.1103/PhysRevE.80.036307 \\
		10    & C2    & 8.35  & $256^{3}$ & Carbonate & http://dx.doi.org/10.1103/PhysRevE.80.036307 \\
		11    & Doddington \#1 & 5.41  & $256^{3}$ & Sandstone & https://doi.org/10.1016/j.advwatres.2012.03.003 \\
		12    & Doddington \#2 & 5.41  & $256^{3}$ & Sandstone & https://doi.org/10.1016/j.advwatres.2012.03.004 \\
		13    & Doddington \#3 & 5.41  & $256^{3}$ & Sandstone & https://doi.org/10.1016/j.advwatres.2012.03.005 \\
		14    & Doddington \#4 & 5.41  & $256^{3}$ & Sandstone & https://doi.org/10.1016/j.advwatres.2012.03.006 \\
		15    & Estaillades \#1 & 10.72 & $256^{3}$ & Carbonate & https://doi.org/10.1016/j.physa.2009.12.006 \\
		16    & Fontainebleau \#1 & 17.17 & $256^{3}$ & Sandstone & https://doi.org/10.1016/j.physa.2009.12.006 \\
		17    & Fontainebleau \#2 & 17.17 & $256^{3}$ & Sandstone & https://doi.org/10.1016/j.physa.2009.12.006 \\
		18    & Fontainebleau \#3 & 17.17 & $256^{3}$ & Sandstone & https://doi.org/10.1016/j.physa.2009.12.006 \\
		19    & Fontainebleau \#4 & 17.17 & $256^{3}$ & Sandstone & https://doi.org/10.1016/j.physa.2009.12.006 \\
		20    & Fontainebleau \#5 & 17.17 & $256^{3}$ & Sandstone & https://doi.org/10.1016/j.physa.2009.12.006 \\
		21    & Fontainebleau \#6 & 17.17 & $256^{3}$ & Sandstone & https://doi.org/10.1016/j.physa.2009.12.006 \\
		22    & Ketton & 10.62 & $256^{3}$ & Carbonate & https://doi.org/10.1016/j.advwatres.2012.03.005 \\
		23    & S1    & 10.18 & $256^{3}$ & Sandstone & http://dx.doi.org/10.1103/PhysRevE.80.036307 \\
		24    & S2    & 5.81  & $256^{3}$ & Sandstone & http://dx.doi.org/10.1103/PhysRevE.80.036307 \\
		25    & S3    & 10.66 & $256^{3}$ & Sandstone & http://dx.doi.org/10.1103/PhysRevE.80.036307 \\
		26    & S4    & 10.50 & $256^{3}$ & Sandstone & http://dx.doi.org/10.1103/PhysRevE.80.036307 \\
		27    & S5    & 4.68  & $256^{3}$ & Sandstone & http://dx.doi.org/10.1103/PhysRevE.80.036307 \\
		28    & S6    & 5.98  & $256^{3}$ & Sandstone & http://dx.doi.org/10.1103/PhysRevE.80.036307 \\
		29    & S7    & 5.63  & $256^{3}$ & Sandstone & http://dx.doi.org/10.1103/PhysRevE.80.036307 \\
		30    & S8    & 5.73  & $256^{3}$ & Sandstone & http://dx.doi.org/10.1103/PhysRevE.80.036307 \\
		31    & S9    & 3.98  & $256^{3}$ & Sandstone & http://dx.doi.org/10.1103/PhysRevE.80.036307 \\
		32    & F42A  & 11.72 & $256^{3}$ & Sandpack & http://dx.doi.org/10.1103/PhysRevE.80.036307 \\
		33    & F42B  & 11.72 & $256^{3}$ & Sandpack & http://dx.doi.org/10.1103/PhysRevE.80.036307 \\
		34    & F42C  & 11.72 & $256^{3}$ & Sandpack & http://dx.doi.org/10.1103/PhysRevE.80.036307 \\
		35    & LV60A & 11.72 & $256^{3}$ & Sandpack & http://dx.doi.org/10.1103/PhysRevE.80.036307 \\
		36    & LV60B & 11.68 & $256^{3}$ & Sandpack & http://dx.doi.org/10.1103/PhysRevE.80.036307 \\
		37    & LV60C & 11.72 & $256^{3}$ & Sandpack & http://dx.doi.org/10.1103/PhysRevE.80.036307 \\
		38    & A1    & 4.51  & $256^{3}$ & Sandpack & http://dx.doi.org/10.1103/PhysRevE.80.036307 \\
		39    & Benthemier \#2 & 4.97  & $256^{3}$ & Sandstone & https://doi.org/10.1016/j.advwatres.2012.03.005 \\
		40    & Benthemier \#3 & 4.97  & $256^{3}$ & Sandstone & https://doi.org/10.1016/j.advwatres.2012.03.005 \\
		41    & Benthemier \#4 & 4.97  & $256^{3}$ & Sandstone & https://doi.org/10.1016/j.advwatres.2012.03.005 \\
		42    & Benthemier \#5 & 4.97  & $256^{3}$ & Sandstone & https://doi.org/10.1016/j.advwatres.2012.03.005 \\
		43    & Benthemier \#6 & 4.97  & $256^{3}$ & Sandstone & https://doi.org/10.1016/j.advwatres.2012.03.005 \\
		44    & Benthemier \#7 & 4.97  & $256^{3}$ & Sandstone & https://doi.org/10.1016/j.advwatres.2012.03.005 \\
		45    & Berea \#8 & 7.11  & $256^{3}$ & Sandstone & https://doi.org/10.1016/j.advwatres.2015.07.012 \\
		46    & Clashach & 10.15 & $256^{3}$ & Sandstone & https://doi.org/10.1016/j.advwatres.2015.07.012 \\
		47    & Doddington \#5 & 9.50  & $256^{3}$ & Sandstone & https://doi.org/10.1016/j.advwatres.2015.07.012 \\
		48    & Estaillades \#2 & 10.16 & $256^{3}$ & Carbonate & https://doi.org/10.1016/j.advwatres.2015.07.012 \\
		49    & Indiana & 10.15 & $256^{3}$ & Carbonate & https://doi.org/10.1016/j.advwatres.2015.07.012 \\
		50    & Ketton \#2 & 10.15 & $256^{3}$ & Carbonate & https://doi.org/10.1016/j.advwatres.2015.07.012 \\
		51    & Monte Gamb. \#1 & 4.94  & $256^{3}$ & Carbonate & https://doi.org/10.1016/j.advwatres.2012.03.004 \\
		52    & Monte Gamb. \#2 & 4.94  & $256^{3}$ & Carbonate & https://doi.org/10.1016/j.advwatres.2012.03.004 \\
		53    & Monte Gamb. \#3 & 4.94  & $256^{3}$ & Carbonate & https://doi.org/10.1016/j.advwatres.2012.03.004 \\
		54    & Monte Gamb. \#4 & 4.94  & $256^{3}$ & Carbonate & https://doi.org/10.1016/j.advwatres.2012.03.004 \\
		55    & Monte Gamb. \#5 & 4.94  & $256^{3}$ & Carbonate & https://doi.org/10.1016/j.advwatres.2012.03.004 \\
		56    & Monte Gamb. \#6 & 4.94  & $256^{3}$ & Carbonate & https://doi.org/10.1016/j.advwatres.2012.03.004 \\
		57    & Monte Gamb. \#7 & 4.94  & $256^{3}$ & Carbonate & https://doi.org/10.1016/j.advwatres.2012.03.004 \\
		58    & Monte Gamb. \#8 & 4.94  & $256^{3}$ & Carbonate & https://doi.org/10.1016/j.advwatres.2012.03.004 \\
		59    & Hollington \#1 & 2.17  & $256^{3}$ & Sandstone & https://doi.org/10.1007/s11242-019-01244-8 \\
		60    & Hollington \#2 & 2.17  & $256^{3}$ & Sandstone & https://doi.org/10.1007/s11242-019-01244-8 \\
		\hline

\caption{Sources of the original tomography data and some specifications including original names, size and spatial resolutions.}
\label{tab:data_source}
\end{longtable}	

	\normalsize

\section*{Appendix B}

\begin{table}[H]
	\footnotesize
	\centering
	\begin{tabular}{c|c| c }
		\hline	
		Feature               & Training--validation (K--S distance) & Training--test  (K--S distance) \\
			\hline
Absolute Perm.             & 0.025 & 0.118 \\
Formation Factor           & 0.013 & 0.019 \\
Cementation Factor         & 0.017 & 0.021 \\
Pore Density               & 0.013 & 0.025 \\
Tortuosity                 & 0.018 & 0.018 \\
Avg. Connectivity          & 0.022 & 0.017 \\
Avg. Throat Rad.           & 0.016 & 0.014 \\
Avg. Pore Rad.             & 0.015 & 0.023 \\
Avg. Throat Length         & 0.017 & 0.021 \\
Avg. Pore Inscribed Rad.   & 0.012 & 0.023 \\
Specific Surface           & 0.024 & 0.020 \\
Avg. Throat Inscribed Rad. & 0.014 & 0.016 \\
Grain Sphericity           & 0.012 & 0.020 \\
Avg. Grain Rad.            & 0.014 & 0.009 \\
Rel. Young Module          & 0.013 & 0.018 \\
		\hline   
	
	\end{tabular}
\smallskip
	\caption{Results of two--sample Kolmogorov--Smirnov (K--S) test to check the similarity/dissimilarity of the training--validation and training--test data, K--S distances close to 1 indicate dissimilarity of the distributions.}
\label{tab:ks_test}

\end{table}

	\newpage

\section*{Bibliography}
\bibliography{ref}

\end{document}